\documentclass[aps,pre,epsf,superscriptaddress,amsmath,amssymb,amsfonts,twocolumn,showpacs]{revtex4-1}

\usepackage[utf8]{inputenc}

\usepackage{amsmath}
\usepackage{amssymb}
\usepackage{amsfonts}
\usepackage{amstext}
\usepackage{amsthm}
\usepackage{mathtools}
\DeclarePairedDelimiter\ceil{\lceil}{\rceil}
\usepackage{physics}

\usepackage{graphicx}

\usepackage{epsfig}
\usepackage{dcolumn}
\usepackage{bm}
\usepackage{braket}
\usepackage{color}
\usepackage[colorlinks=true,citecolor=blue]{hyperref}

\usepackage[english]{babel}

\usepackage{multirow}

\allowdisplaybreaks
\advance\parskip 0.4pt

\usepackage{natbib}

\begin{document}

\title{ On-demand  generation of dark soliton trains in Bose-Einstein condensates}
\author{A. Romero-Ros}
\affiliation{Center for Optical Quantum Technologies, Department of Physics,
University of Hamburg, Luruper Chaussee 149, 22761 Hamburg,
Germany}

\author{G. C. Katsimiga}
\affiliation{Center for Optical Quantum Technologies, Department of Physics,
University of Hamburg, Luruper Chaussee 149, 22761 Hamburg,
Germany}

\author{P. G. Kevrekidis}
\affiliation{Department of Mathematics and Statistics, University
of Massachusetts Amherst, Amherst, MA 01003-4515, USA}

\author{B. Prinari}
\affiliation{Department of Mathematics and Department of Physics, State University
of New York, Buffalo, New York 14260, USA}

\author{G. Biondini}
\affiliation{Department of Mathematics and Department of Physics, State University
of New York, Buffalo, New York 14260, USA}

\author{P. Schmelcher}
\affiliation{Center for Optical Quantum Technologies, Department of Physics,
University of Hamburg, Luruper Chaussee 149, 22761 Hamburg,
Germany}\affiliation{The Hamburg Centre for Ultrafast Imaging,
University of Hamburg, Luruper Chaussee 149, 22761 Hamburg,
Germany}

\date{\today}

\begin{abstract}
Matter-wave interference mechanisms  in  one-dimensional Bose-Einstein condensates  that  allow for the controlled generation of dark soliton trains upon choosing suitable box-type initial configurations are described.
First, the direct  scattering problem for the defocusing nonlinear Schrödinger equation with nonzero boundary conditions  and general box-type initial configurations is discussed, and expressions for the discrete spectrum corresponding to the dark soliton excitations generated by the dynamics are obtained. 
It is found that the size of the initial box directly affects the number, size and velocity of the solitons, while the initial phase determines the parity of the solutions.
The analytical results are compared to those of numerical simulations of the Gross-Pitaevskii equation, both in the absence and in the presence of a harmonic trap. 
The numerical results  bear out the analytical results with excellent agreement.
\end{abstract}

\maketitle

\section{Introduction} \label{sec:introduction}

Dark solitons are fundamental nonlinear excitations stemming from the balance between dispersion and  suitable kinds of  nonlinearity.
They are found to arise in diverse physical systems ranging from water waves~\cite{Chabchoub2013} and magnetic materials~\cite{Tong2010} to nonlinear optics~\cite{Zakharov1973,Corney1997,Kivshar1998}
and Bose-Einstein condensates (BECs)~\cite{Pethick2008,Pitaevskii2016,Becker2008,Frantzeskakis2010}.
For instance, in nonlinear optics dark solitons emerge in media with positive dispersion and defocusing nonlinearity whose evolution is described by the so-called defocusing nonlinear Schr{\"o}dinger (NLS) equation~\cite{Kevrekidis2015}.
On the other hand,  in the BEC context  dark solitons  form in systems with repulsive interatomic interactions~\cite{Kevrekidis2007} obeying the so-called Gross-Pitaevskii equation (GPE).

BECs, due to their high degree of controllability and isolation from the
environment~\cite{Bloch2008},
constitute fertile physical platforms for investigating the existence, dynamics and
interactions~\cite{Huang2001,Stellmer2008,Kamchatnov2009,Jezek2016}
of these matter-waves or multi-component~\cite{Hoefer2011,Yan2012,Bersano2018}
and multi-dimensional variants thereof~\cite{Denschlag2000,Anderson2001,Shomroni2009}.
Additionally, several powerful techniques have been utilized in order to generate
such waves. These include, among others, phase imprint~\cite{Burger1999,Denschlag2000,Becker2008} and density engineering~\cite{Shomroni2009}, perturbing the BEC with localized impurities~\cite{Dutton2001,Engels2007} and interference experiments~\cite{Reinhardt1997,Scott1998,Weller2008,Theocharis2010}.

Among  the aforementioned methods, the latter is based on the matter-wave interference of two colliding condensates, a process via which dark soliton trains can be produced.
Several experimental and theoretical works have been devoted  to  studying the controllable creation of such dark soliton arrays~\cite{Reinhardt1997,Scott1998,Weller2008,Brazhnyi2003,Romero-Ros2019}.
They revealed, among  other things,  that the momenta of the colliding BEC parts and their
relative phase play an important role in the number of generated solitonic entities.
This result  has been derived analytically for the defocusing NLS equation by means of the inverse scattering transform (IST) in the seminal work of Ref.~\cite{Zakharov1973}.
Recent theoretical attempts have exploited the integrable nature of the above scalar NLS model and further developed an IST formalism accounting for both symmetric~\cite{Demontis2013,Biondini2014} and fully asymmetric non-zero-boundary conditions (NZBC)~\cite{Biondini2016a}.

In the present work we exploit the unprecedented level of control that the ultracold environment offers along with the exact analytical tools provided by both direct scattering methods and the IST with NZBC and we report the  on-demand  generation of dark soliton arrays.
In particular,  we consider  a one-dimensional (1D), harmonically trapped scalar BEC composed of repulsively interacting atoms,  and we study the response of such a system to box-type initial  configurations ~\cite{Zakharov1973,Espinola-Rocha2009,Biondini2014,Romero-Ros2019}
(see also Ref.~\cite{Gredeskul1989,Swartzlander1991,Ostrovskaya1999} in nonlinear optics)
whose shape is controlled by five distinct parameters.
Limiting cases of the latter directly mimic interference and density/phase engineering processes
suggesting the experimental relevance of our findings.
The closest analogue to this in the context of trapped BECs that we are familiar with appears in the work of~\cite{Brazhnyi2003}, however that work is based on the (approximate) Bohr-Sommerfeld
quantization rule for hyperbolic function based perturbations of
the initial density or  phase profile. Here, on the other hand, we leverage both the
pioneering work of~\cite{Zakharov1973} and also the recent developments
of~\cite{Demontis2013,Biondini2014}, to obtain explicit analytical
results based on IST and then extend them via suitable approximations
in the trapped case. 

More specifically,  first we consider the integrable version of the problem,
i.e., the defocusing NLS equation with NZBC.
The direct scattering problem for this equation with the above box-type initial condition is solved analytically.
Expressions for the discrete eigenvalues of the  scattering problem, which as usual  determine the amplitudes and the velocities of the ensuing dark solitons,  are found, and the exact soliton waveforms and the center of each of them can be extracted within the IST. 
Having at hand the exact analytical expressions, a systematic study of the dynamical evolution of the scalar system is  then  put forth.
Distinct parameter explorations are conducted including, for instance, in-phase (IP) and out-of-phase (OP) initial  configurations.  
In all cases investigated herein, remarkable agreement between the analytical predictions and our numerical findings is observed.
This agreement in turn means that,  for example,  the number of dark solitons that are expected to nucleate via interference is a-priori predicted by our initial condition, along with the amplitudes and velocities of the emergent matter waves.
It is also found that the size of the initial  box  directly affects the number, the amplitude and velocity of the emitted dark  solitons.
Additionally, its phase,  which can be now manipulated with the analytical tools  discussed in this work,  along with its depth can determine not only the even or odd number of nucleated dark solitons, but can also lead to an asymmetrical distribution thereof.
Remarkably, the analytical predictions can be suitably extended in the presence of a harmonic confinement.
Specifically, it is found that  in each scenario, besides the anticipated modifications in the amplitudes and velocities of the emitted dark solitons, stemming from confinement, the general behavior of the trapped system closely follows that of the homogeneous setting  (where by ``homogeneous'' we mean the case without confinement). 
Additionally here, by monitoring during evolution the center of mass of each nucleated dark soliton, estimations of the velocities, the amplitudes and finally the oscillation frequency of individual waves are obtained.
Excellent agreement with the analytical expressions is exposed for the soliton amplitudes and velocities, while deviations smaller than $4\%$ are identified for the oscillation frequency when compared to the analytical predictions~\cite{Frantzeskakis2010}.

The  flow  of this  paper  is as follows.
In  Section~\ref{sec:model} we introduce the model and discuss the direct scattering problem for the NLS with a general box-type   initial condition.
Additionally, we comment on limiting cases, in terms of the involved  box  parameters, and thus establish  connections with interference and density/phase engineering processes used in contemporary BEC experiments.
In  Section~\ref{sec:results} we present our findings.
First, we extract the eigenvalues of the scattering problem over a wide range of different initial configurations. 
Then,  we perform  a comparison of the analytical predictions with direct numerical simulations of the GPE both in the absence and in the presence of the trap.
Finally, in  Section~\ref{sec:conclusions} we summarize our results and discuss possible  directions for future study.

\section{Model setup, scattering problem and discrete eigenvalues} 
\label{sec:model}

\subsection{ The Gross-Pitaevskii and nonlinear Schr\"odinger equation setup }

The system of interest is a scalar 1D BEC consisting of repulsively interacting atoms
being confined in a highly anisotropic trap with longitudinal and transverse
trapping frequencies chosen such that $\omega_x\ll \omega_{\perp}$.
In such a cigar shaped geometry~\cite{Becker2008,Hoefer2011}, the condensate wavefunction along
the transverse direction, being the ground state of the respective harmonic oscillator,
can be integrated out. 
Then, in the mean-field framework, the BEC dynamics for the longitudinal part of the wavefunction $\Psi(x,t)$ is governed by the following 1D GPE~\cite{Pethick2008,Pitaevskii2016}
\begin{equation}
    i\hbar\frac{\partial\Psi}{\partial t} = - \frac{\hbar^2}{2m}\frac{\partial^2\Psi}{\partial x^2} + V(x)\Psi+g|\Psi|^2\Psi.
    \label{eq:GPEdim}
\end{equation}
Moreover, in the above expression $V(x)=m\omega_x^2 x^2/2$ denotes the external harmonic potential.
Additionally, $m$ denotes the atomic mass, while $g=2\hbar\omega_{\perp}a_s$ is the effective 1D coupling constant expressed in terms of the $s$-wave scattering length, $a_s$.
The latter accounts for two-atom collisions and can be tuned by means of Feshbach resonances~\cite{Inouye1998,Chin2010}.
In the present work we consider $g=1$ and our setup can be realized experimentally by considering e.g. a gas of $^{87}$Rb atoms~\cite{Pethick2008,Pitaevskii2016}.
By performing the transformations: $|q|^2=2a_s|\Psi|^2$, $x'=a^{-1}_\perp x$, with $a_{\perp}=\sqrt{\hbar/m\omega_\perp}$ being the transverse oscillator length, and $t'=\omega_\perp t$,
we cast the aforementioned scalar GPE in the dimensionless form
\begin{equation}
    i\frac{\partial q}{\partial t} = -\frac{1}{2} \frac{\partial^2q}{\partial x^2} + \frac{1}{2}\Omega^2x^2 q+|q|^2 q\,,
    \label{eq:GPE}
\end{equation}
where  $\Omega\equiv\omega_x/\omega_{\perp}$.
For convenience we further dropped the primes.
In the absence of a trapping potential (i.e., for $\Omega=0$), Eq.~\eqref{eq:GPE} 
reduces to the well-known defocusing NLS equation~\cite{Kevrekidis2015}.

The latter integrable model can be solved analytically via IST and it is known to possess
dark soliton solutions that have NZBC at infinity~\cite{Biondini2014}.
To this end for the analytical considerations to be carried out below, we further perform the rescaling $\Tilde{q}(x,t)=q(\sqrt{2} x,t)\exp{-2i q_o^2t}$  in the integrable version of Eq.~(\ref{eq:GPE}) and by omitting the tildes
we end up with
\begin{align}
    iq_t+q_{xx}-2(\abs{q}^2-q_o^2)q=0.
    \label{eq:manakov_system}
\end{align}
Notice that with the aforementioned transformation Eq.~(\ref{eq:manakov_system}) satisfies the following time-independent NZBC at infinity
\begin{align}
    \lim_{x\to\pm\infty} q(x,t)=q_\pm=q_oe^{i\theta_\pm}.
\end{align}
Henceforth, $q_o=\abs{q_o}>0$ (without loss of generality), $\theta_\pm$ are real numbers, and the subscripts $t$ and $x$ introduced in Eq.~(\ref{eq:manakov_system}) denote here and throughout this work partial differentiation with respect to time and space, respectively.

\begin{figure}[t]
    \centering
    \includegraphics[width=\columnwidth]{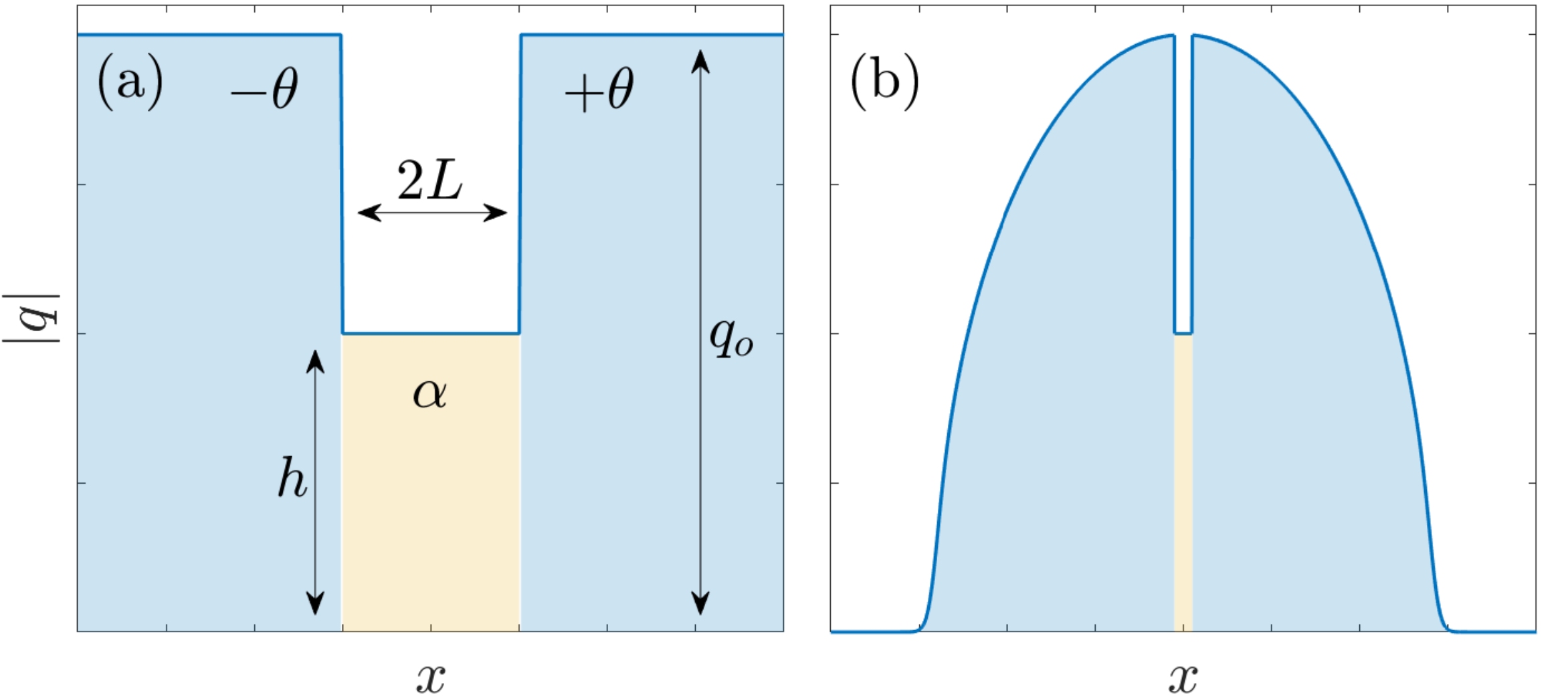}
    \caption{Schematic illustration of the box-type initial configuration~\eqref{eq:initial_conditions},
    for  generic  wavefunction parameters, i.e., $L,\,q_o,\, \theta,\, h$ and $\alpha$ (a) in the absence and (b) in the presence of a harmonic trapping potential.
    }
    \label{fig:IC}
\end{figure}

Motivated by our recent work of Ref.~\cite{Romero-Ros2019}, but also by the earlier works of Refs.~\cite{Zakharov1973,Krokel1988,Gredeskul1989,Ostrovskaya1999,Kamchatnov2002,Nikolov2004,
Dabrowska-Wuster2009} regarding the controllable nucleation of soliton arrays, for our analytical and
numerical investigations below, we  consider the following box-type initial configurations  for the condensate wavefunction:
\begin{align}
    q(x,0) =
\left\{
	\begin{array}{ll}
		q_oe^{i\theta_-},   & \quad x < -L\,,\\
		he^{i\alpha},    & \quad \abs{x} < L\,,\\
		q_oe^{i\theta_+},   & \quad x > L\,.
	\end{array}
\right.
\label{eq:initial_conditions}
\end{align}
Here, $h \geq 0$ refers to the depth ($h<q_o$) or height ($h>q_o$) of the box.
Additionally, $L$ corresponds to the half width of the  box,  $q_o$ is  the background  amplitude, $\theta_\pm$ are the  asymptotic phases at either side of the box  and $\alpha$ is  the phase inside the box.
It will be convenient to introduce the quantities
\begin{equation}
\Delta\theta=\theta_+-\theta_-\,,\quad
\Delta\theta_+=\theta_+-\alpha\,,\quad
\Delta\theta_-=\alpha-\theta_-\,, 
\end{equation}
to denote the distinct phase differences in each of the different regions of the box.
A schematic illustration of the initial  configuration~\eqref{eq:initial_conditions}  is provided in Fig.~\ref{fig:IC}(a).
Owing to the phase invariance of the NLS equation, we can take $\theta_+ = - \theta_- = \theta$
without loss of generality, and we will do so hereafter.
We will refer to the cases $\Delta\theta= 0$ and $\Delta\theta \neq 0$ as in-phase (IP) and out-of-phase (OP) condensates, respectively, and to the special case $h=0$ (which describes the complete absence of atoms inside the box) as that of a ``zero box''.

\subsection{ Direct scattering of box-type configurations} 
\label{sec:scattering_problem}

Here, we follow the presentation of~\cite{Biondini2014}.
As noted earlier, the defocusing NLS equation [Eq.~\eqref{eq:manakov_system}] is  an integrable nonlinear partial differential equation, for which many initial value problems  can be solved by means of the IST  via its  Lax pair.
The $2\times 2$ Lax pair associated with Eq.~\eqref{eq:manakov_system} is
\begin{align}
    \bm{\phi}_x=\vb{X}\bm{\phi}\,, \qquad \bm{\phi}_t=\vb{T}\bm{\phi}\,,
    \label{eq:scattering _problem}
\end{align}
where $\bm\phi$ is a $2\times 2$ matrix eigenvector,
\begin{eqnarray}
\vb{X}(x,t,k)&=&ik\vb{J}+\vb{Q}\,, \\
\vb{T}(x,t,k)&=&2ik^2\vb{J}-i\vb{J}(\vb{Q}_x-\vb{Q}+q_o^2)-2k\vb{Q}\,,
    \label{eq:X_T}
\end{eqnarray}
and 
\begin{gather}
    \vb{J}=\mqty(-1 & 0 \\ 0 & 1)\,, \qquad
    \vb{Q}(x,t)=\mqty(0 & q \\ q^* & 0) \,.
    \label{eq:J_Q}
\end{gather}
The first equation in \eqref{eq:scattering _problem} is referred to as the scattering problem, $k \in \mathbb{C}$ as the scattering parameter, and $q(x,t)$ as the scattering potential.
One can expect that, as $x\to\pm\infty$, the solutions of the direct scattering problem are approximated by those of the asymptotic scattering problem $\bm{\phi}_x=\vb{X}_\pm\bm{\phi}$, where $\vb{X}_\pm=-ik\vb{J}+\vb{Q}_\pm$ and $\vb{Q}_\pm=\lim_{x\to\pm\infty}\vb{Q}(x,t)$.

The eigenvalues of $\vb{X}_{\pm}$ are $\pm i\lambda$, where
\begin{align}
    \lambda(k) = \sqrt{k^2-q_o^2} \,.
\end{align}
As in Refs.~\cite{Prinari2006,Biondini2014,Biondini2015,Biondini2015a}, we take the branch cut along the semilines $(\infty,-q_o)$ and $(q_o,\infty)$, and  we define uniquely $\lambda(k)$ by requiring  that $\Im\lambda(k)\geq0$.
(This corresponds to working on one sheet of the two-sheeted Riemann surface
defined by $\lambda(k)$~\cite{Prinari2006,Biondini2014,Biondini2015,Biondini2015a}).

Here, we define the Jost solutions $\bm{\phi}_\pm(x,t,k)$ as the simultaneous solutions of both parts of the Lax pair satisfying the boundary conditions
\begin{align}
    \bm{\phi}_\pm(x,t,k)\equiv \vb{Y}_\pm(k)e^{i\vb{\Theta}(x,t,k)} + {\mathcal{O}}(1) \qq{as} x\to\pm\infty \,
    \label{eq:jost_solution}
\end{align}
where $\vb{\Theta}(x,t,k)=\vb{\Lambda}x-\vb{\Omega}t$, $\vb{\Lambda}=\text{diag}(-\lambda,\lambda)$, $\vb{\Omega}=\text{diag}(2k\lambda, -2k\lambda)$, and $\vb{Y}_\pm(k)$ are the simultaneous eigenvector matrices of $\vb{X}_\pm$ and $\vb{T}_\pm$. 
Both Jost solutions are related to each other through the scattering relation
\begin{align}
    \bm{\phi}_-(x,t,k) = \bm{\phi}_+(x,t,k) \vb{S}(k) \,,
\end{align}
and the scattering coefficients (the entries of the $2\times 2$ scattering matrix $S(k)$) are time independent on account of the fact that the Jost eigenfunctions are chosen to be simultaneous solutions of the Lax pair.

As we are only concerned with the discrete eigenvalues of the scattering operator,
which are time-independent, hereafter we will consider the scattering problem at $t = 0$ and omit the time dependence from the
eigenfunctions.
At $t=0$ the scattering problem in each of the three regions $x<-L$, $|x|<L$, and $x>L$ takes the form $v_x=(-ik\vb{J}+\vb{Q}_j)v$
with $j=c,\pm$ with constant potentials $\vb{Q}_\pm$ and $\vb{Q}_c$,
\begin{align}
    \vb{Q}_{\pm} = \mqty( 0 & q_oe^{\pm i\theta} \\
                    q_oe^{\mp i\theta} & 0 ) \,,
    \;\;
    \vb{Q}_c = \mqty( 0 & he^{i\alpha} \\
                    he^{-i\alpha} & 0 ) \,,
    \label{eq:Q}
\end{align}
where  again we set  $\theta_+=-\theta_-=\theta$ without loss of generality.
One can then easily find explicit solutions for the scattering problem in each of the three regions:
\begin{subequations}
    \label{eq:jost_eigenfunctions}
    \begin{align}
        \bm{\varphi}_l(x,k) &= \vb{Y}_-(k)e^{i\vb{\Lambda} x} \qquad x\le -L \\
        \bm{\varphi}_c(x,k) &= \vb{Y}_c(k)e^{i\vb{M} x} \qquad |x|\le L \\
        \bm{\varphi}_r(x,k) &= \vb{Y}_+(k)e^{i\vb{\Lambda} x} \qquad x\ge L
    \end{align}
\end{subequations}
where $\vb{M}=\text{diag}(-\mu,\mu)$, with $\mu=\sqrt{k^2-h^2}$, and
\begin{eqnarray}
\vb{Y}_{\pm}(k) =& \mqty( k+\lambda & -iq_oe^{\pm i\theta} \\
                             iq_oe^{\mp i\theta} & k+\lambda ), \\ 
    \vb{Y}_c(k) =& \mqty( k+\mu & -ihe^{i\alpha} \\
                         ihe^{-i\alpha} & k+\mu ). 
    \label{eq:Y}
\end{eqnarray}
We then have explicit representations for the Jost solutions $\bm{\phi}_\pm(x,0,k)$ in their respective regions, i.e. $\bm{\phi}_{-}(x,0,k)\equiv \bm{\varphi}_l(x,k)$ for $x \leq -L$, and $\bm{\phi}_{+}(x,0,k)\equiv \bm{\varphi}_r(x,k)$ for $x\geq L$.
At the boundary of each region one can express the fundamental solution on the left as a linear combination of the fundamental solution on the right, and vice versa.
In particular, we can introduce scattering matrices $S_{-}(k)$ and $S_{+}(k)$ such that
\begin{subequations}
    \label{eq:parcial_scattering_matrices}
    \begin{align}
        \bm{\varphi}_- (-L,k) &= \bm{\varphi}_c(-L,k) \vb{S}_- (k) \,, \\
        \bm{\varphi}_c (L,k) &= \bm{\varphi}_+(L,k) \vb{S}_+ (k) \,.
    \end{align}
\end{subequations}
As a consequence, we can express the scattering matrix $S(k)$ relating the Jost solutions $\bm{\phi}_\pm(x, k)$ as
\begin{align}
    \vb{S}(k) &= \vb{S}_+(k)\vb{S}_-(k) \nonumber \\
              &= e^{-i\vb{\Lambda} L} \vb{Y}_+^{-1}\vb{Y}_c
                 e^{2i\vb{M}L}\vb{Y}_c^{-1}\vb{Y}_-e^{-i\vb{\Lambda} L} \,.
        \label{eq:scatterting_matrix}
\end{align}
Computing the right-hand side of Eq.~\eqref{eq:scatterting_matrix}, we obtain the following expression
for the first element $s_{11}(k)$ of the scattering matrix $\vb{S}(k)$:
\begin{eqnarray}
\lambda\mu e^{-i(2\lambda L+\theta)}s_{11}(k) &=& \mu\cos(2\mu L)\left(\lambda\cos\theta-ik\sin\theta\right) \nonumber \\
    &+&i\sin(2\mu L)\Big[hq_o\cos\alpha \nonumber \\
    &-&k(k\cos\theta-i\lambda\sin\theta)\Big].
    \label{eq:s11}
\end{eqnarray}
The discrete eigenvalues of the scattering problem are the zeros of $s_{11}(k)$.
Each of them contributes a dark soliton to the solution.
For the scalar defocusing NLS equation the zeros are real and simple, and there is a finite number of them, belonging to the spectral gap $k\in(-q_o,q_o)$~\cite{Faddeev2007}.
In the case of a single zero $k_o$, the dark soliton solution of Eq.~(\ref{eq:manakov_system}) reads 
\begin{eqnarray}
q_d(x,t)&=& q_o\cos\beta_o -iq_o\sin\beta_o \times \nonumber \\
&\times & \tanh\Big[\sin\beta_o \left(x-x_0+q_o\cos\beta_o t\right)\Big],  
    \label{eq:dark_soliton}
\end{eqnarray}
where $k_o=q_o\cos\beta_o$, and $\lambda_o=iq_o\sin\beta_o$ provide the velocity and the amplitude of the soliton,
\begin{subequations}
    \label{eq:dark_soliton_parameters}
    \begin{align}
        v &= -q_o\cos\beta_o \equiv -k_o \,, \\
        A_d &= q_o\sin\beta_o \equiv \sqrt{q_o^2-k_o^2} \,,
    \end{align}
\end{subequations}
respectively, and $x_0$ stands for the center of the soliton.

We  point out that  the maximum soliton speed $|v_\mathrm{ max }|=q_o$,  which coincides with the speed of sound of the condensate, $c=q_o$ (note that $c=\sqrt{gn}$~\cite{Bogoliubov1947,Lee1957} in the dimensionless units adopted herein, with $n$ being the density of the BEC).
Recall (cf. Eq.~\eqref{eq:dark_soliton_parameters}) that a true soliton can never reach such speed ($v=k_o<q_o$).
On the other hand, the maximum amplitude of a soliton is $A_d^\mathrm{ max }=q_o$, and it is attained by solitons with $v=k_o=0$, also known as black solitons.
In what follows, we will use  the variable  $k_o$ to refer to a generic zero or to a set of zeros.

\subsection{ Special cases, symmetries , interference and phase/density engineering} \label{sec:analogies}

Some of the most popular methods to generate dark solitons in 1D BECs are phase imprinting, density engineering and colliding condensates, as discussed in the introduction.
In this section we show how  box-type initial configurations  can be analogous to most setups
used  in  the aforementioned methods  for  the generation of dark solitons in 1D BECs, and we obtain analytical  results  in the corresponding cases.

Before  discussing  each case, it is worth  noting that, regardless of  the method of creation,  configurations with a phase difference  $\Delta\theta = \pi$ allow the emergence of black solitons.
Recall that black solitons are static solitons, i.e. $v=k_o=0$ [see Eq.~\eqref{eq:dark_soliton_parameters}].
We can establish straightforward  necessary and sufficient conditions to ensure that $k = 0$ is a discrete eigenvalue, i.e., a zero of $s_{11}(k)$.
Since we are looking for zeros, from now on it is convenient to  only work with the right-hand side of Eq.~\eqref{eq:s11}.
When  $k = 0$ both $\lambda$ and $\mu$ are purely imaginary, i.e. $\lambda = iq_o$ and $\mu=ih$, and  Eq.~\eqref{eq:s11} yields 
\begin{align}
    \cosh(2hL)\cos\theta+\sinh(2hL)\cos\alpha  =0 \,.
    \label{eq:s11_k_0}
\end{align}
Thus, $k = 0$ is a discrete eigenvalue if and only if either
(i) $\cos\theta = \cos\alpha = 0$, for any choice of $h, L , q_o$;
or (ii) $\tanh(2hL)=-\cos\theta\sec\alpha$.
The former is in line with the previous statement regarding black solitons, i.e., $\theta=\pi/2$.
The latter obviously requires $\cos\theta\sec\alpha>-1$.

Equation~\eqref{eq:s11_k_0} is a special case of the symmetries possessed by the discrete spectrum in certain configurations. 
Since $\lambda$ and $\mu$ are both even functions of $k$,
when $\theta=0$ (corresponding to an in-phase background, i.e., $\Delta\theta=0$),
the right-hand side of Eq.~\eqref{eq:s11} is also an even function of $k$. 
Thus, independently of the value of $h$ and $\alpha$,
to each discrete eigenvalue $k_o\ne0$ there corresponds a symmetric discrete eigenvalue $-k_o$,
yielding a pair of symmetric solitons with the same amplitude and opposite velocity. 
The same symmetry also arises when $\theta = \pi/2$ (i.e., $\Delta\theta=\pi$)
if either $h=0$ or $\alpha = \pi/2$, since in this case $s_{11}(k)$ becomes an odd function of $k$.

We now discuss how the box-type configurations~\eqref{eq:initial_conditions} relate to two of the aforementioned methods, associated with the interference process.
Such a setup in principle consists of two condensates, e.g. of the same atomic species, being separated from each other by some distance.
The emergence of dark solitons in this setting relies on matter-wave interference phenomena
occurring during the collision of the condensates~\cite{Reinhardt1997,Scott1998,Weller2008,Theocharis2010}.
Basically, when the condensates collide an interference pattern appears.
Then, depending on the initial momenta and phase of the colliding condensates, some of the interference fringes formed might develop into dark solitons.
Specifically, the number of the latter is known to be proportional to the momenta of the colliding
condensates~\cite{Scott1998,Weller2008} and can be increased by placing them farther apart.
Additionally, also known is that the parity of the number of solitons depends on the phase difference between the condensates.
Namely, an even (odd) number of them is going to emerge if the initial condensates are IP (OP).
A  box-type initial configuration  that can mimic such an interference process is that with $h=0$.
In this case the two sides of the box represent the two independent colliding condensates, being separated by a distance $2L$ and having a phase difference $\Delta\theta=2\theta$.
Taking $h=0$, Eq.~\eqref{eq:s11} reduces to
\begin{eqnarray}
0 &=& k\cos\left(2kL\right)\Big[\lambda\cos\theta-ik\sin\theta \Big] \nonumber \\
      &-& ik\sin \left(2kL\right)\Big[k\cos\theta-i\lambda\sin\theta\Big], 
    \label{eq:s11_h_0}
\end{eqnarray}
which can be rewritten as
\begin{align}
    \sqrt{k^2-q_o^2}k\cos(2k L+\theta)-ik^2\sin(2k L+\theta) = 0 \,.
\end{align}
Apart from the trivial solution $k=0$, the other solutions $k_n$ are given by
\begin{align}
    2k_nL+\theta = \text{arctan}\qty(\frac{\sqrt{q_o^2-k_n^2}}{k_n})+\pi n  \,,
    \label{eq:s11_h_0_kn}
\end{align}
with $n\in\mathbb{Z}$ (note: in Sec.~\ref{sec:results} the index $n$ is replaced by $o$).
This sets all the solutions in the interval $-q_o<k_n<q_o$, as expected.
Moreover, the limiting case of $k_{n} \to q_o$ provides the number of zeros $\mathcal{N}$ for a given $L$
and $0\leq\theta\leq\pi$ as
\begin{align}
    \mathcal{N} = \ceil[\bigg]{\frac{2q_o L + \theta}{\pi}}.
    \label{eq:num_solutions}
\end{align}
In the above expression $\ceil{\;}$ denotes the ceiling function (Eq.~\eqref{eq:num_solutions} was already derived in~\cite{Zakharov1973,Espinola-Rocha2009}).
From the above equation, it is then clear that the number of solitons (zeros) is proportional to the distance between the colliding condensates, and its parity depends on their phase difference.

We now discuss the second methodology,  namely phase-imprinting~\cite{Dobrek1999,Burger1999,Becker2008}.
This technique imprints a phase-jump on the condensate, by exposing part of it to a far-detuned laser beam, which can dynamically develop into dark solitons.	
This setting can be reproduced by  box-type initial configurations even with $L=0$.  This extreme case represents the setting of a highly localized in space phase imprinting.  
Notice that such a choice indeed leads to a condensate that has two regions with different phases.
Then, Eq.~\eqref{eq:s11} directly reduces to
\begin{align}
    \lambda\cos{\theta} - ik\sin{\theta} = 0 \,,
\end{align}
which yields a single zero
\begin{align}
    k = q_o\cos{\theta}=q_o\cos\qty(\frac{\Delta\theta}{2}).
    \label{eq:s11_L_0}
\end{align}
Notice also that a black soliton solution occurs when $\Delta\theta=\pi$, as expected from Eq.~\eqref{eq:s11_k_0}(i).
Even though Eq.~\eqref{eq:s11_L_0}, having a single phase-jump, does not produce soliton trains,
it nevertheless assures the controlled generation of a single soliton given a particular $\theta$.
Moreover it correctly captures earlier findings~\cite{Wu2002,Becker2008,Stellmer2008,Fritsch2020}
according to which the generated solitons are faster, shallower and wider,
the smaller the phase difference is [see also Eq.~\eqref{eq:dark_soliton_parameters}].

We can consider other cases as well. For instance a case in which a phase is imprinted on a finite region of the BEC,resulting in a three-section condensate with two phase-jumps~\cite{Wu2002,Becker2008,Stellmer2008,Fritsch2020}.
To reproduce such a setup with the  box-type initial configuration of Eq.~(\ref{eq:initial_conditions}) we consider a homogeneous condensate $(h=q_o)$ having an extent $2L$ and to which we impose a phase $(\alpha)$.
In this case, if $L\approx q_o$ then Eq.~\eqref{eq:s11} needs to be numerically solved.
Yet, in the limit $L\gg q_o$, we can treat both phase-jumps as being far apart from each other to treat them locally.
Thereby, we can make use again of Eq.~\eqref{eq:s11_L_0} with the appropriate phase difference
\begin{align}
    k_\pm = q_o\cos\qty(\frac{\Delta\theta_\pm}{2})= q_o\cos\qty(\frac{\theta\mp\alpha}{2})\,.
    \label{eq:s11_h_q_L_0}
\end{align}
Recall that $\pm$ denotes the right or the left phase-jump [see Eq.~\eqref{eq:initial_conditions}].
Here, we want to point out that $0\leq\Delta\theta_\pm\leq\pi$, otherwise it needs to be transformed accordingly with a $\pi$ shift.
Yet, another case example consists of a  box-type initial configuration corresponding to  a barrier on top of a background, i.e., $h>q_o$.
Considering $h\gg q_o$ and imposing a phase $\alpha$ at the location of the barrier, Eq.~\eqref{eq:s11} can be expressed as
\begin{align}
    \tanh(2hL) = \frac{k\sin\theta-\sqrt{q_o^2-k^2}\cos\theta}{q_o\cos\alpha} \,.
    \label{eq:s11_h_gg_q}
\end{align}
Since the left-hand side is always positive, Eq.~\eqref{eq:s11_h_gg_q} provides zeros if and only if $\theta$ and $\alpha$ are such that they produce a positive right-hand side.
For example, if we look for zeros corresponding to black solitons, i.e. $k_o=0$,
we recover condition (ii) from Eq.~\eqref{eq:s11_k_0}.
Additionally, in the limit $L, h\to\infty$, Eq.~\eqref{eq:s11_h_gg_q} reduces to
\begin{align}
    k_\pm= q_o\sin\Delta\theta_\pm \,.
    \label{eq:s11_h_inf}
\end{align}

Lastly, we briefly comment on the analogy of density engineering methods with our  box-type initial configurations. 
These methods are typically used to create density defects on a condensate, which can be small~\cite{Dutton2001} or substantial~\cite{Engels2007} depletions of the latter.
To mimic such techniques  with  our  box-type initial configurations,  each case needs to be considered individually and the zeros must be found numerically by solving Eq.~\eqref{eq:s11}.
Specific case examples of the zeros and their parametric dependencies for distinct  box-type initial configurations  are presented in the forthcoming section.

\section{Dark soliton generation and dynamics} \label{sec:results}

\subsection{Analytical  results for the discrete spectrum} 
\label{sec:sol_Analytical}

Here we analytically characterize the dark solitons produced by the box-type initial configurations~\eqref{eq:initial_conditions} by studying  the zeros of the first element, $s_{11}(k)$, of the scattering matrix, $\vb{S}(k)$ [Eq.~\eqref{eq:s11}], upon considering different selections of the system parameters.
Specifically, we utilize the wavefunction of Eq.~(\ref{eq:initial_conditions})
which is characterized by the following five parameters:
the half width, $L$, the amplitude, $q_o$, the side phase, $\pm\theta$,
the depth (or height) of the box, $h$, and its phase, $\alpha$ [see also Fig.~\ref{fig:IC}(a)].
To sort out all the spectra, we choose a set defined by two main variables,
that  will be  varied while the remaining system parameters are held fixed.
Since $L$ and $h$ can be thought of as the main parameters of the scalar system under consideration,
 the following  discussion will be mainly focused on the set  of values of $L$ and $h$. 
The corresponding exploration, in terms of parametric variations, is performed for the following selection of the configuration parameters:
\begin{equation}
    L\in[1,9],\;\; \theta = \qty{0,\frac{\pi}{2}},\;\; h\in[0,q_o],\;\; \alpha=\qty{0,\pi} \,,
\end{equation}
together with $q_o =1$.
However, we will also briefly comment on other selections too whose results are not included herein
for brevity.

In what follows, we present the spectra of zeros of the first element $s_{11}(k)$ of the scattering matrix for three different sets  of values of $L$ and $h$. 
Each distinct exploration is shown in a figure consisting of ten panels (a) to (i) that range from $L=1$ to $L=9$, respectively.
Each panel contains different zeros, $k_o$, as $h$ is varied, with each of which corresponding to a particular dark soliton solution.

\begin{figure}[t]
    \centering
    \includegraphics[width=0.9\columnwidth]{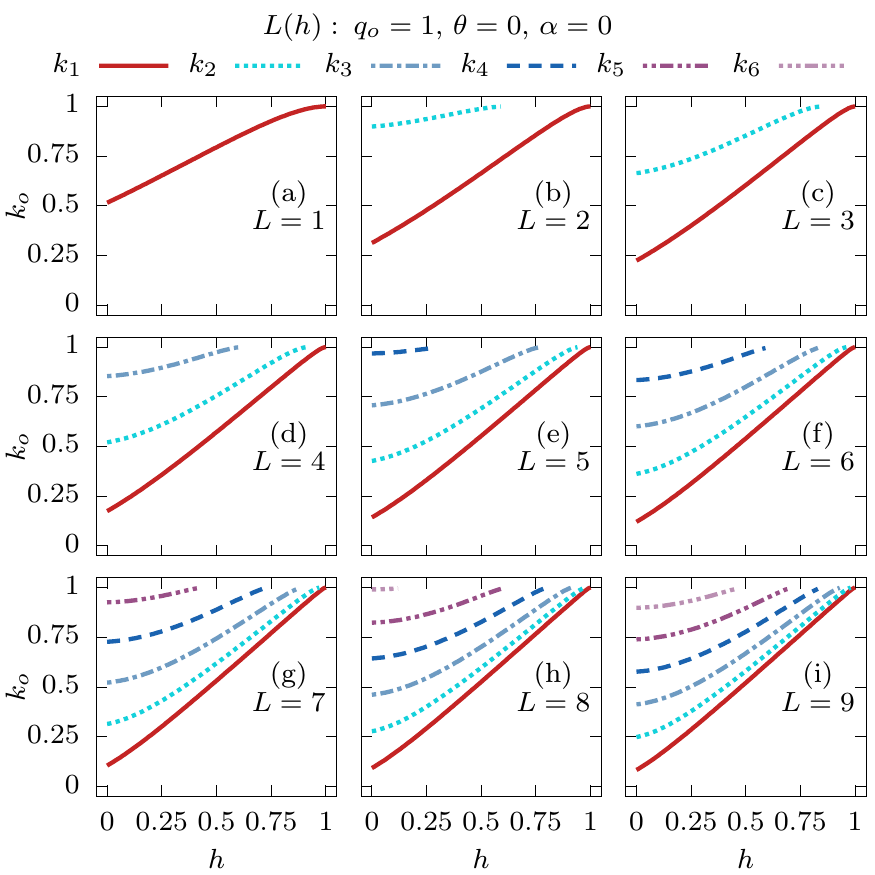}
    \caption{Zeros of $s_{11}(k)$ as a function of $h$ for different values of $L$.
    The parameters $q_o=1$, $\theta=0$ and $\alpha=0$ remain fixed.
    Only $k_o>0$ are shown due to the parity of the zeros.}
    \label{fig:roots_Lh_q_1_a_0_th_0}
\end{figure}
\smallskip
\textbf{ All in-phase.}
The first selection  we investigate is the case  $q_o=1$, $\theta=0$ and $\alpha=0$.
Here, $\Delta\theta=0$ implying an IP  configuration  [see Fig.~\ref{fig:IC}(a)],
and $\alpha=0$ implies that the box is also in-phase with the background.
The corresponding  spectra of zeros is presented in Figs.~\ref{fig:roots_Lh_q_1_a_0_th_0}(a)--(i).
Due to the parity of the zeros, only $k_o>0$ are shown in the aforementioned figure.
As can be directly seen, increasing $L$ increases the number of solitons (i.e. the number of $k_o$'s).
Particularly, when $L=1$ only one pair of zeros, $k_1$, appears (one-pair of soliton solutions)
while $L=5$ [$L=9$] allows up to four [six] pairs of them, $k_1,\dots,k_4$ [$k_1,\dots,k_6$], to occur.
This is in agreement with the analytical expression of Eq.~\eqref{eq:num_solutions} and correctly captures the $h=0$ case.
Recall that $h=0$  is referred to as a ``zero box'' and is physically associated with a setting of independent condensates colliding. 
Note also that even though Eq.~\eqref{eq:num_solutions} is not a general expression but rather a limiting case, the number of solitons still remains proportional to $L$ and $q_o$ even when $h\neq 0$.
Also by inspecting Figs.~\ref{fig:roots_Lh_q_1_a_0_th_0}(a)--(i), it becomes apparent that
for fixed $h$, increasing $L$ decreases the value of $k_o$.
 This implies that the resulting solitons are slower as $L$
increases [see Eq.~\eqref{eq:dark_soliton_parameters}], which can be understood as the momenta available in the system being distributed among a larger number of solitons.
This trend can be easily discerned by monitoring e.g. $k_1(h=0)$ as $L$ increases [see also
Eq.~\eqref{eq:s11_h_0_kn}].
Indeed, initially, i.e. for $L=1$, $k_1(h=0)=0.515$ [Fig.~\ref{fig:roots_Lh_q_1_a_0_th_0}(a)].
Then, for $L=2$, $k_1$ decreases to $k_1(h=0)=0.313$ [Fig.~\ref{fig:roots_Lh_q_1_a_0_th_0}(b)]
and already for $L=9$ $k_1(h=0)=0.083$ [Fig.~\ref{fig:roots_Lh_q_1_a_0_th_0}(i)].
On the other hand,  for a fixed $L$ it is found that the value of $k_o$ increases,  i.e., the solitons become faster, upon increasing $h$.
Moreover, since $k\in(-q_o,q_o)$ [see also Sec.~\ref{sec:model}], this increasing tendency of $k_o$ for increasing $h$ holds as such until $k_o=q_o$, a threshold above which solitons cease to exist [see Eq.~\eqref{eq:dark_soliton_parameters}].
Recalling now that increasing $h$ implies that the initial jump in the configuration becomes progressively shallower,
then when $h=q_o$ there is no box configuration that can lead to the creation of solitonic excitations.
Such outcome also persists  for $h>q_o$.

\begin{figure}[t]
    \centering
    \includegraphics[width=0.9\columnwidth]{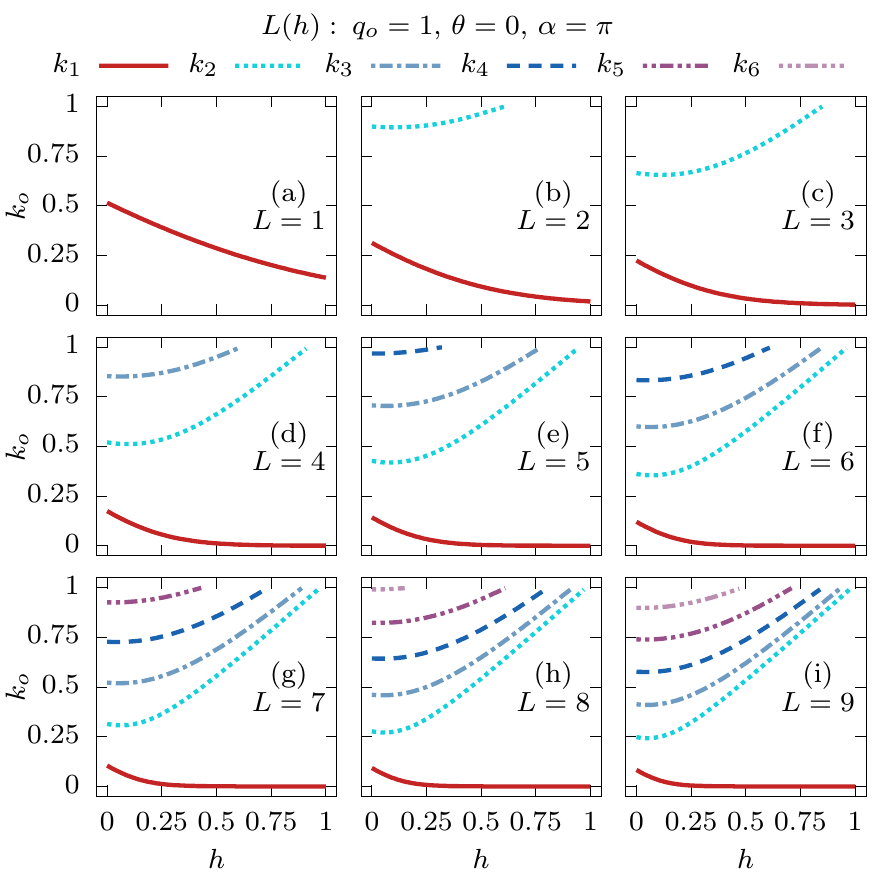}
    \caption{Zeros of $s_{11}(k)$ as a function of $h$ for different values of $L$.
    The parameters $q_o=1$, $\theta=0$ and $\alpha=\pi$ remain fixed.
    Only $k_o>0$ are shown due to the parity of the zeros.}
    \label{fig:roots_Lh_q_1_a_pi_th_0}
\end{figure}
\smallskip
\textbf{ Out-of-phase box.}
Next we turn to the second selection of parameters, in which $q_o=1$, $\theta=0$  as before, but where now $\alpha=\pi$.
This is also an IP configuration, but the box is now out-of-phase with the background.
The analytical solutions, given by the zeros of the first scattering element,
are illustrated in Figs.~\ref{fig:roots_Lh_q_1_a_pi_th_0}(a)--(i).
Since  $\Delta\theta=0$ here as well, we only show the range $k_o>0$, as before. 
Below, we solely focus on $k_1$ since it is the only zero having a distinct trend when compared to those shown in Fig.~\ref{fig:roots_Lh_q_1_a_0_th_0}.
Notice that contrary to the aforementioned zeros, and also to the previous parameter selection,
as $h$ increases $k_1$ decreases  with the associated soliton
thus becoming slower  and, in fact, $k_1 \rightarrow 0$ as $h
\rightarrow \infty$.
This decreasing tendency of $k_1$ is in agreement with Eq.~\eqref{eq:s11_k_0} and specifically with condition (ii).

Additionally, it is also evident  from  Figs.~\ref{fig:roots_Lh_q_1_a_pi_th_0}(b)--(i) that $k_1\to 0$ as $L\to\infty$ independently of $h$.
A discrete eigenvalue $k_1=0$ would in theory correspond to a pair of black solitons,  each  generated as a consequence of the phase jump $\Delta\theta_\pm=\mp\pi$ at $x=\pm L$.
In turn, this would correspond to $k_1=0$ being a degenerate eigenvalue with degeneracy two.
However, it is well-known that, for the scalar defocusing NLS, all discrete eigenvalues are simple~\cite{Faddeev2007}, and no coalescence of zeros is possible, in contrast to the focusing case.
What is happening is that, as $L\to\infty$, one reaches an approximate degeneracy:
when the phase jumps at $x=\pm L$ are sufficiently far apart from each other, one can approximately treat them as independent scattering problems.
Then, the solution to each problem is simply given by Eq.~\eqref{eq:s11_L_0}, which indeed coincides with the  observed  result.
Nonetheless, it is important to realize that the discrete eigenvalues of the overall system are only approximately given by those of the individual scattering problems, and a careful analytical treatment shows that in practice the symmetric pair of discrete eigenvalues is always at a nonzero distance from $k=0$, although this distance vanishes in the limit $L\to\infty$.

Finally, we note in passing that cases corresponding to different choices of $\alpha$ have also been explored, for which upon increasing $h$, $k_1\rightarrow k_{\pm}$ [Eq.~\eqref{eq:s11_h_q_L_0}].
To be precise, it is found that if $0\leq\alpha\leq\pi/2$ then $k_1$ increases and eventually reaches $k_1=q_o$.
On the other hand, if $\pi/2<\alpha\leq\pi$, then $k_1$ asymptotically tends to a different yet again finite value,  as $h$ is increased.
Indeed, taking the limit $h\gg q_o$ and for $\theta=0$,  Eq.~\eqref{eq:s11_h_gg_q} yields
\begin{align}
    \tanh(2hL)=-\sec\alpha\sqrt{1-\left(\frac{k}{q_o}\right)^2} \,,
\end{align}
which directly implies that $\sec\alpha<0$ explaining this way that there exist
values of $\alpha$ for which $k_1=q_o$ is reached.
Past this point, and for $L\to\infty$ or $h\to\infty$, $k_1\to q_o\sin\alpha$
asymptotically slow [see Eq.~(\ref{eq:s11_h_inf})].

\begin{figure}[t]
    \centering
    \includegraphics[width=0.9\columnwidth]{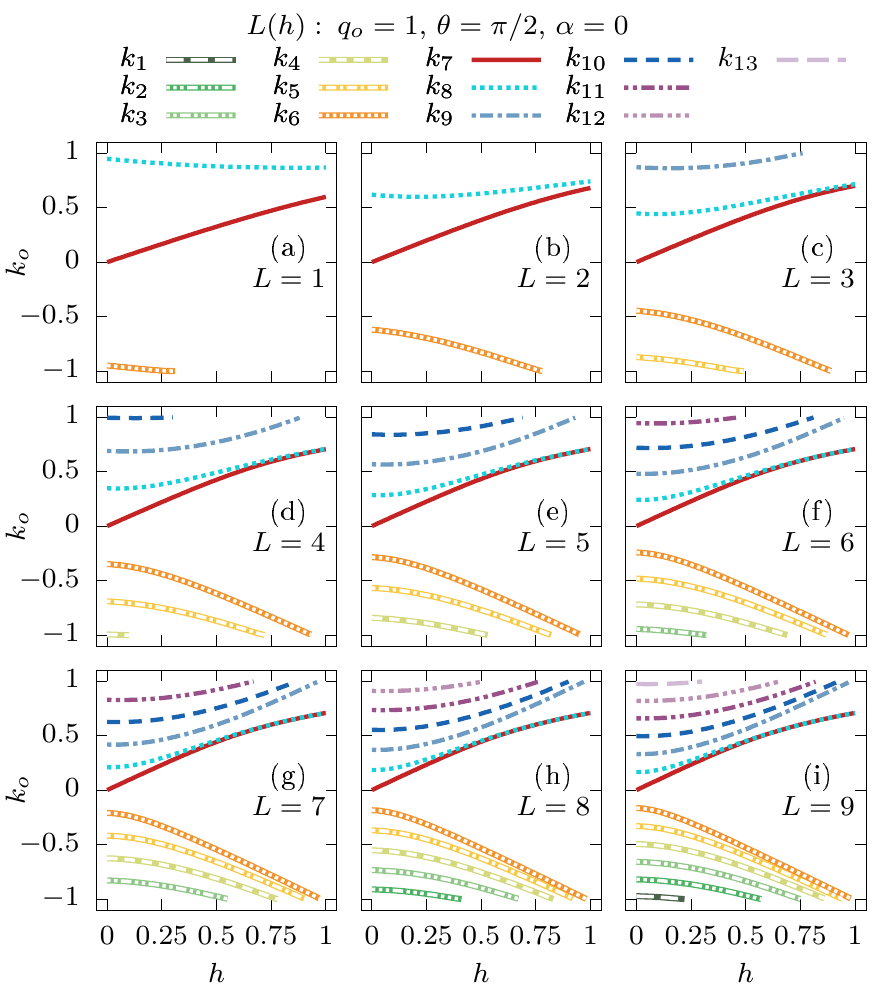}
    \caption{Zeros of $s_{11}(k)$ as a function of $h$ for different values of $L$.
    The parameters $q_o=1$, $\theta=\pi/2$ and $\alpha=0$ remain fixed.}
    \label{fig:roots_Lh_q_1_a_0_th_pi2}
\end{figure}
\smallskip
\textbf{ Asymptotic phase difference.}
Our  last parametric exploration, shown in Figs.~\ref{fig:roots_Lh_q_1_a_0_th_pi2}(a)--(i),
consists of various choices of $L$ and $h$ as before, but with  the remaining system parameters as
$q_o=1$, $\theta=\pi/2$ and $\alpha=0$.
This initial state preparation corresponds to an OP box-type configuration, with $\Delta\theta=\pi$.
In contrast to the previous cases, this choice produces an asymmetric distribution of  discrete eigenvalues. 
This outcome is evident by looking at the zeros as $h$ is varied, as illustrated in Figs.~\ref{fig:roots_Lh_q_1_a_0_th_pi2}(a)--(i).
Exceptionally, for $h=0$ all zeros are paired, i.e. $k_o=\pm k$, except for the $k_7$ one.
For instance, for $L=9$ and $h=0$ thirteen soliton solutions are identified, corresponding
to the thirteen distinct zeros, $k_1,\ldots,k_{13}$, shown in Fig.~\ref{fig:roots_Lh_q_1_a_0_th_pi2}(i).
From these, solutions $k_1,\ldots,k_6=-k_{13},\ldots,-k_8$, respectively.
As in the preceding  scenarios, it is clear that  in the present case the number of solitons also increases as $L$ increases, and increasing $L$ while keeping $h$ fixed results in zeros that have
smaller $|k_o|$ value  and are thus slower. 
Additionally, for fixed $L$ the number of expected soliton solutions decreases as we increase $h$.
For example, for $L=3$ all five solutions $k_5,\ldots,k_9$ occur e.g. at $h=0$, but only four of them, i.e. $k_6,\ldots,k_9$, are left for $h=0.6$,  further  reducing to three ($k_6, k_7$ and $k_8$) for $h=0.8$ [Fig.~\ref{fig:roots_Lh_q_1_a_0_th_pi2}(c)].
Moreover, increasing $h$ produces also an increase in the magnitude of each zero ($|k_o|$)
until eventually $|k_o|=q_o$ is reached, leading in turn to the absence of soliton solutions.

Exceptions to the aforementioned general behavior of the solutions are the zeros $k_7$ and $k_8$ that never  reach the threshold  $|k_o|=q_o$ for $h\leq q_o$.
Instead, these two solutions are seen to merge asymptotically as $h$ increases, a merging that occurs faster for larger $L$ values.
This merging can in turn be translated into two (asymptotically) identical solitons, having the same velocity and amplitude, but different soliton centers, $x_0$ [see Eqs.~(\ref{eq:dark_soliton_parameters})].
To understand further the aforementioned behavior, we considered also different values of $\theta$ which in turn unraveled that if $h=q_o$ then $k_7= k_8 \to q_o\cos(\theta/2)$ as $L \to \infty$ [see Eq.~\eqref{eq:s11_h_q_L_0}].
This is also in line with our interpretation for the existence of degenerate zeros in the scalar NLS
(see also our previous discussion).
On the other hand, if $h\to\infty$ then $k_7=k_8\to q_o\sin(\Delta\theta_\pm)$ independently of $L$ [see Eq.~\eqref{eq:s11_h_inf}].
Note here that the subscripts referring to the solutions $k_7, k_8$ are such for the specific case example addressed herein.
However, different values of $\theta$ might change the number of solutions and thus their relevant labelling.

\subsection{Nucleation of dark soliton trains:  Without confinement} \label{sec:sol_GPE}

In this section  we aim to validate the analytical results presented in Section~\ref{sec:sol_Analytical} (and more specifically to bear out the discrete eigenvalues identified there) by numerically solving the scalar GPE in the absence of a confining potential, i.e. $\Omega=0$ [Eq.~(\ref{eq:GPE})].
For the dynamical evolution of the aforementioned scalar system, we employ a fourth-order Runge-Kutta integrator accompanied by a second order finite-difference method that accounts for the spatial derivatives.
The spatial and temporal discretizations introduced are $\dd x=0.1$ and $\dd t=0.001$, respectively, and the position of the boundaries used in the dynamics is at $|x|=2500$ to avoid finite size effects.
In the following, we fix $L=5$ and $q_o=1$ and we consider as representative examples the values $h=\{0, 0.5\}$.
Additionally, for this $h$ selection, we further consider the cases of $\theta=\qty{0,\pi/2}$ and $\alpha=\qty{0,\pi}$.

Below we present our findings regarding the dynamical nucleation of dark solitons via the matter-wave interference of two colliding condensates~\cite{Weller2008,Reinhardt1997,Scott1998,Theocharis2010}
for various initial configurations.
When comparing the analytical predictions to the numerical observations, it is important to keep in mind that the  various solitons generated by the initial conditions~\eqref{eq:initial_conditions} 
are in general interacting with each other. 
Therefore, one can expect to be able to visually identify individual solitons only in the asymptotic limit of $x \to \pm\infty$,  after the solitons emerge from the creation process and can be considered to be  well-separated and independent from one another.
 Conversely, during  the initial stages of the dynamics  one expects to see discrepancies  between
the analytically  determined solitons  and the numerically  observed  ones.
One can also expect any such discrepancies to become smaller and gradually disappear as $t\to\pm\infty$.
This expectation is indeed reflected by the results, as discussed below.

\begin{figure}[t]
    \centering
    \includegraphics[width=\columnwidth]{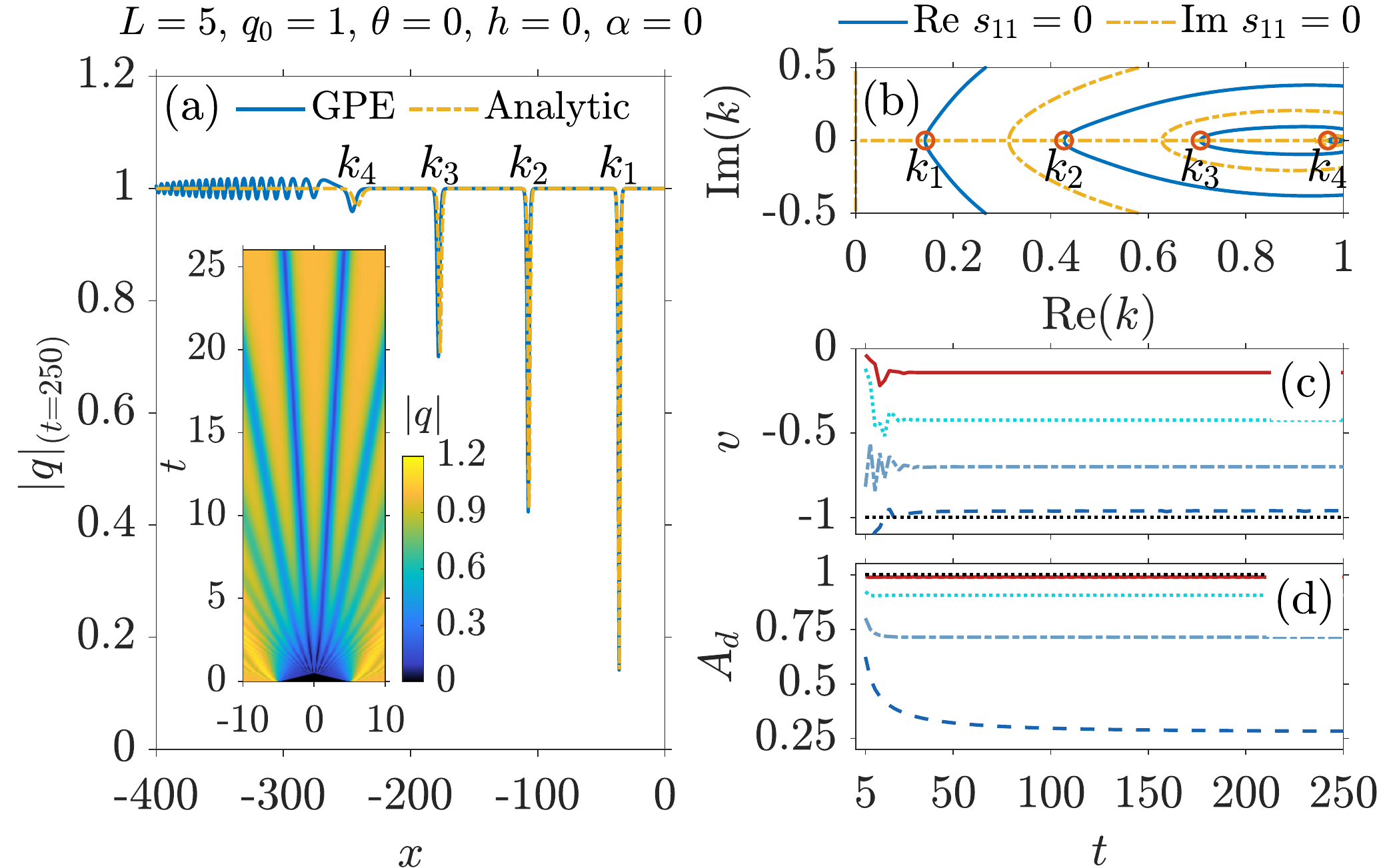}
    \caption{Dark soliton solutions resulting from the
    box-type initial condition~\eqref{eq:initial_conditions}
    with $L=5$, $q_o=1$, $\theta=0$, $h=0$ and $\alpha=0$,
    corresponding to a zero box and an in-phase background
    [cf. Fig.~\ref{fig:roots_Lh_q_1_a_0_th_0}(e)].
    (a) Snapshot of $\abs{q}$ at $t=250$ given by the GPE (solid blue line) and the analytical
    solutions with $x_0(t=0)=0$ (dashed yellow line).
    The inset shows the spatiotemporal evolution of $\abs{q}$ at initial times.
    (b) Contour plot of $\Re s_{11}=0$ (solid blue line) and $\Im s_{11}=0$ (dashed yellow line) on the complex $k$-plane for  $\Re k \ge 0$. 
    The zeros, $k_o$, are depicted by red circles.
    Temporal evolution of  the velocities (c) and the amplitudes (d)  of the dark solitons.
    In both (c) and (d) the distinct lines (from bottom to top) correspond to the analytical predictions
    stemming from the zeros (from right to left) in (b).
    Dotted black line in (c) refers to the speed of sound and in (d) to the maximum amplitude.
    The zeros, $k_o$, follow the notation introduced in the legend of Fig.~\ref{fig:roots_Lh_q_1_a_0_th_0}, with
    $k_1 = 0.1428$, $k_2 = 0.4271$, $k_3 = 0.7069$ and $k_4 = 0.9608 $.}
    \label{fig:h_0_a_0_th_0}
\end{figure}

\smallskip
\textbf{ Zero box, in-phase background.}
 We start presenting our  findings in Figs.~\ref{fig:h_0_a_0_th_0}(a)--(d).
According to our analytical estimates [see Fig.~\ref{fig:roots_Lh_q_1_a_0_th_0}(e)
and Eq.~\eqref{eq:num_solutions}] four pairs of dark solitons are expected and indeed form when a  zero box  ($h=0$) IP ($\Delta\theta=0$)  configuration  is utilized.
Note that due to the symmetric nucleation of the matter-waves only the solitons located at $x<0$, having negative velocities, $v<0$, and thus corresponding to the positive zeros, $k_o>0$, occurring at $\Re k \ge 0$  are shown in Figs.~\ref{fig:h_0_a_0_th_0}(a) and \ref{fig:h_0_a_0_th_0}(b).
Remarkable agreement between the analytical solutions and the dynamically nucleated matter-waves is observed already at times $t=250$ during evolution, as illustrated in this profile snapshot of the norm of the wavefunction $|q|_{t=250}$ [Fig.~\ref{fig:h_0_a_0_th_0}(a)].
Notice how the emergent dark solitons spread outwards at their initial stages of formation, i.e., right after the collision of the two sides of the initial  box  around $x=0$.
Such spreading at early times $t<5$, as depicted in the spatiotemporal evolution of $|q|$ [inset of Fig.~\ref{fig:h_0_a_0_th_0}(a)], bends the trajectories of the solitons that are symmetrically emitted around the origin.
However, already at $t\approx25$, where also the trajectories of the propagating solitons become linear, the instantaneous velocities, $v=dx_{CM}/dt$ (see below), of the individual coherent structures reach the asymptotic analytical predictions stemming from the zeros, $k_o$, identified in Fig.~\ref{fig:h_0_a_0_th_0}(b), remaining thereafter nearly constant for all times [Fig.~\ref{fig:h_0_a_0_th_0}(c)].
The same trend holds also for the amplitudes, $A_d$, of the emergent entities illustrated in Fig.~\ref{fig:h_0_a_0_th_0}(d).
Note also that in both Figs.~\ref{fig:h_0_a_0_th_0}(c) and \ref{fig:h_0_a_0_th_0}(d) the fastest dark wave denoted by $k_4$ has a velocity proximal to the speed of sound $c=q_0=1$ [dotted black line in Fig.~\ref{fig:h_0_a_0_th_0}(c)], while the slowest soliton denoted by $k_1$ has an amplitude close to the maximum one, i.e., $A_{d}^\mathrm{ max }=q_0=1$ [dotted black line in Fig.~\ref{fig:h_0_a_0_th_0}(d)].

Finally, it is important to mention at this point that, in order to obtain the amplitude of each of the aforementioned solitons (and also for the cases to be presented below), we numerically followed the dark soliton minima during evolution.
Then, the amplitude corresponds to the value of $|q|$ at these minima.
For measuring the instantaneous velocity, we used instead the position given by the center of mass, i.e., $x_{CM} = \left(\int_{x_l}^{x_r}x|q|^2\text{d}x\right)/\left(\int_{x_l}^{x_r}|q|^2\text{d}x\right)$, of each soliton with ${x_{l,r}}$ denoting the area of integration around each dark soliton's core.
Therefore, at early times, the oscillations observed in the temporal evolution of $v$ [Fig.~\ref{fig:h_0_a_0_th_0}(c)] stem from the discrepancies in the calculation of $x_{CM}$.
Indeed, at the initial stages of the dynamics, the calculation of $x_{CM}$ might present some irregular oscillations if a soliton is not well formed nor separated enough from its neighbours or the emitted radiation.
The latter, seen for instance at $x<-275$ in Fig.~\ref{fig:h_0_a_0_th_0}, is a direct effect of the highly excited initial state introduced herein.

\begin{figure}[t]
    \centering
    \includegraphics[width=\columnwidth]{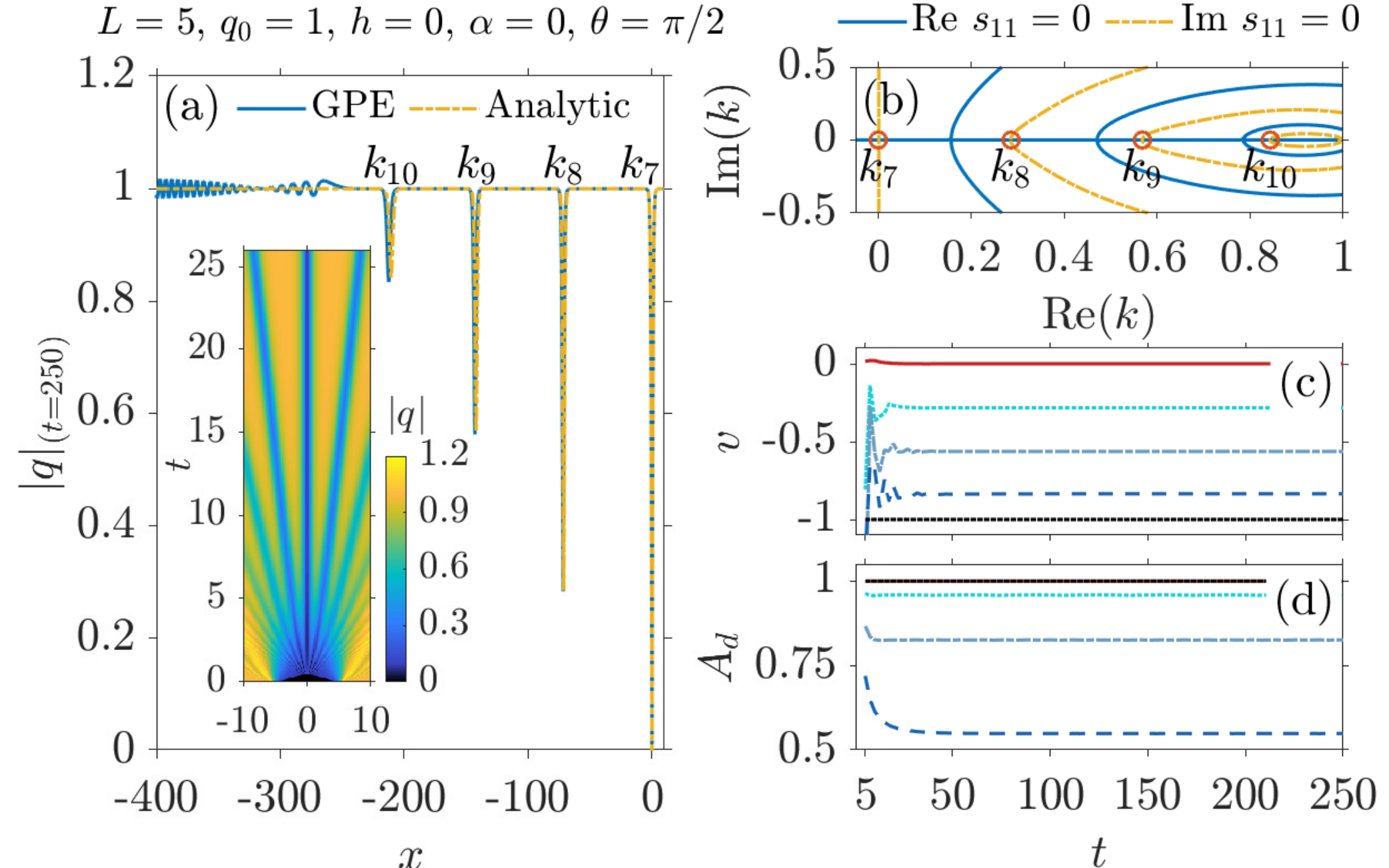}
    \caption{Same as Fig.~\ref{fig:h_0_a_0_th_0} but for $L=5$, $q_o=1$, $\theta=\pi/2$, $h=0$ and $\alpha=0$,
    corresponding to a zero box and an out-of-phase background.
    From left to right the zeros, $k_o$, in (b) that lead to the solitons formed in (a) are located at: $k_7 = 0.0, k_8 = 0.2852, k_9 = 0.5679, k_{10} = 0.8423$ [see the legends in Fig.~\ref{fig:roots_Lh_q_1_a_0_th_pi2}].}
    \label{fig:h_0_a_0_th_pi2}
\end{figure}
\textbf{ Zero box, out-of-phase background.}
Next we turn to the exploration of the  dynamics upon considering a  zero box but with  OP ($\Delta\theta=\pi$)  background. 
Here, our analytical findings suggest the emergence of an odd number of solitons [see Fig.~\ref{fig:roots_Lh_q_1_a_0_th_pi2}(e) at $h=0$ and Eq.~\eqref{eq:num_solutions}].
 This  outcome is dynamically confirmed  by  Figs.~\ref{fig:h_0_a_0_th_pi2}(a)--(d), which show  
three pairs of dark solitons  being  nucleated  together  with a central black  soliton, adding  up to the expected odd number [Fig.~\ref{fig:h_0_a_0_th_pi2}(a)].
Since once more the generation is symmetric with respect to the origin, only the left moving matter-waves are shown in the snapshot of $|q|_{t=250}$ in Fig.~\ref{fig:h_0_a_0_th_pi2}(a)
that correspond to the zeros $k_7-k_{10}$ illustrated in Fig.~\ref{fig:h_0_a_0_th_pi2}(b).
Notice the close similarities  between  this process  and the previous  one. 
Indeed, besides  the number of nucleated waves, the only discernible difference at the early stages of soliton formation is the generation of the central black soliton [inset Fig.~\ref{fig:h_0_a_0_th_pi2}(a)].
The velocities and amplitudes of the evolved solitons  also  follow a trend analogous to the IP case with minor differences  for  the relevant magnitudes of $v$ and $A_d$ for each individual dark soliton [Fig.~\ref{fig:h_0_a_0_th_pi2}(c) and \ref{fig:h_0_a_0_th_pi2}(d)].
The black soliton ($k_7$) has, as expected, $v=0$ and also the maximum amplitude $A_{d}^\mathrm{ max }=q_0=1$.

\begin{figure}[t]
    \centering
    \includegraphics[width=\columnwidth]{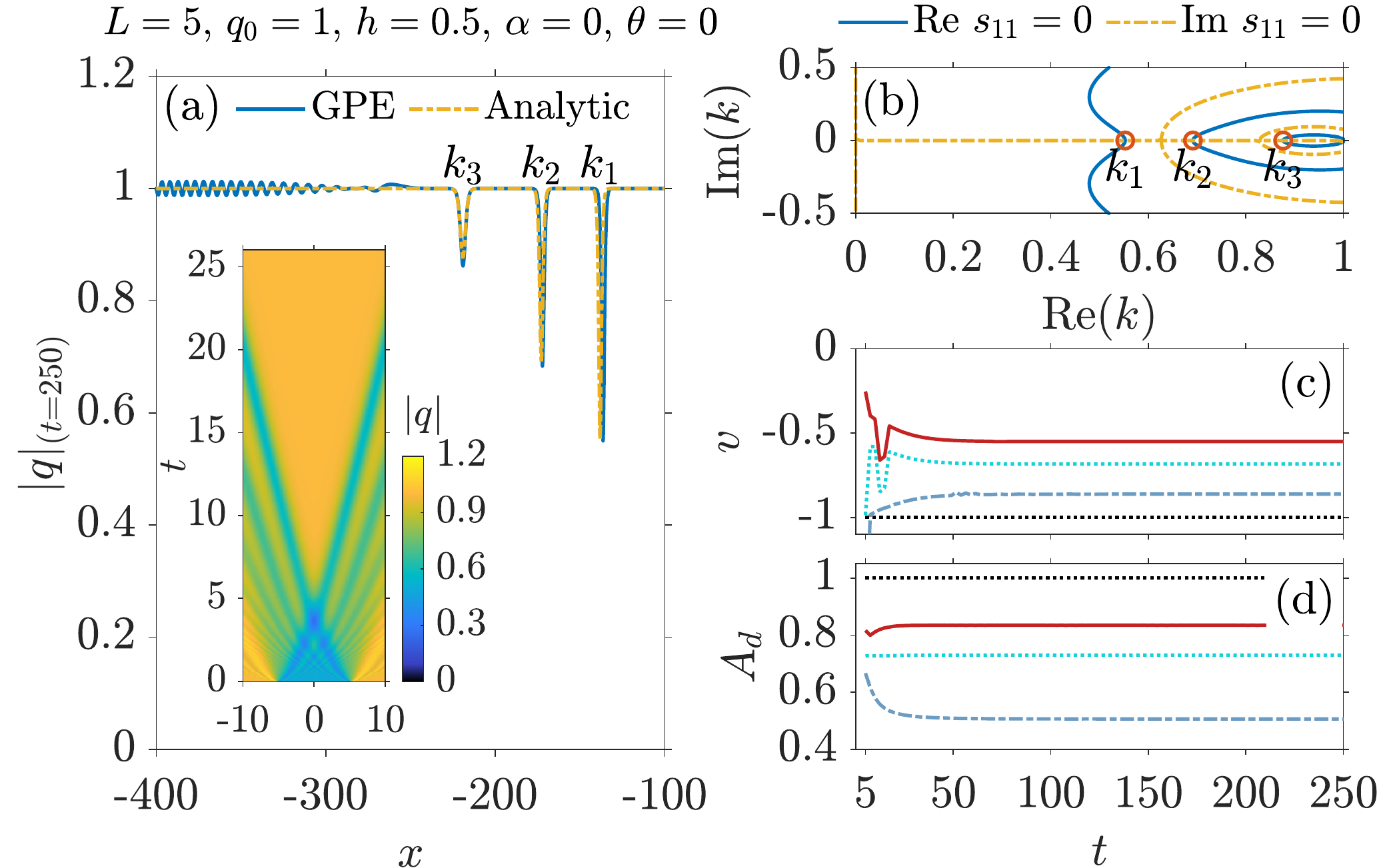}
    \caption{Same as Fig.~\ref{fig:h_0_a_0_th_0} but for $L=5$, $q_o=1$, $\theta=0$, $h=0.5$ and $\alpha=0$, corresponding to a non-zero box in-phase with respect to its background.
    From left to right the zeros, $k_o$, in (b) that lead to the solitons formed in (a) are located at: $k_1 = 0.5526, k_2 = 0.6914, k_3 = 0.8763$ [see the legends in Fig.~\ref{fig:roots_Lh_q_1_a_0_th_0}].}
    \label{fig:h_05_a_0_th_0}
\end{figure}

\begin{figure}[t]
    \centering
    \includegraphics[width=\columnwidth]{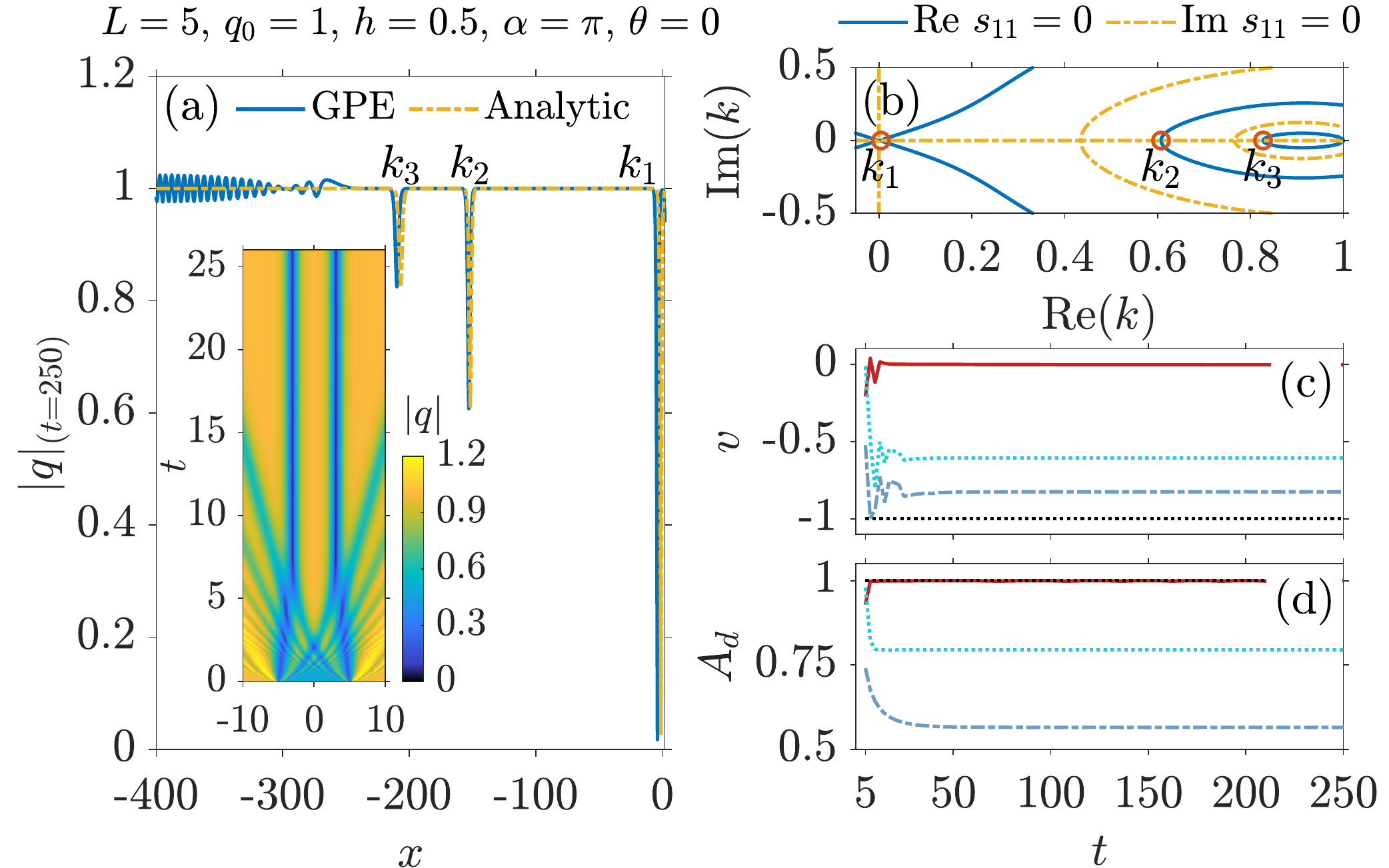}
    \caption{Same as Fig.~\ref{fig:h_0_a_0_th_0} but for $L=5$, $q_o=1$, $\theta=0$, $h=0.5$ and $\alpha=\pi$, corresponding to a non-zero box out-of-phase with respect to its  background.
    From left to right the zeros, $k_o$, in (b) that lead to the solitons formed in (a) are located at: $k_1 = 0.0045, k_2 = 0.06073, k_3 = 0.8269$ [see the legends in Fig.~\ref{fig:roots_Lh_q_1_a_pi_th_0}].}
    \label{fig:h_05_a_pi_th_0}
\end{figure}

\smallskip
\textbf{ Non-zero boxes, dispersive shock waves.}
We  now discuss initial configurations 
whose shape resembles a density defect immersed in the BEC~\cite{Dutton2001,Engels2007,Burger1999}.
To achieve the latter we fix $h=0.5$.
Figures~\ref{fig:h_05_a_0_th_0}(a)--(d) and Figs.~\ref{fig:h_05_a_pi_th_0}(a)--(d)
illustrate representative examples of the dynamical evolution of the scalar system
for IP initial  configurations  but with $\alpha=0$ and $\alpha=\pi$, respectively [see also Fig.~\ref{fig:roots_Lh_q_1_a_0_th_0}(e) and Fig.~\ref{fig:roots_Lh_q_1_a_pi_th_0}(e), respectively].
In both cases, at the initial stages of the dynamics, $t<5$, multiple interference
events significantly distort the homogeneous background and also
disturb the nucleation process.

It should be noted how, in this case as well as the following two, the time evolution generates dispersive shock waves \cite{El2016} as a result of the initial discontinuities.
This is a well-known phenomenon, and in marked contrast to the case when the amplitude in the central box is zero, in which no such structures are generated \cite{Kodama1995}.
The formation and initial dynamics of these dispersive shock waves can be effectively described using Whitham's modulation theory for the defocusing NLS equation 
\cite{Whitham1974,Pavlov1987,Gurevich1987,El1995,Hoefer2006,Hoefer2008}.
In situations where more than one dispersive shock wave is generated, as in the present case
(where each discontinuity generates a separate structure), their interactions can also be effectively studied, as in Refs.~\cite{Biondini2006,Hoefer2007}, using the Whitham modulation equations of higher genus \cite{Forest1986}.
It is also interesting to note that one could still choose to look at the individual oscillations in these dispersive shocks as the initial manifestations of the dark solitons that are the main object of our study.
Also note, however, that the initial speeds of propagation of these individual excitations are quite different from those predicted by the IST, and are instead very well in agreement with the predictions from Whitham modulation theory.
Nonetheless, after these structures have interacted, the final state of the system does become a collection of solitons whose properties agree very well with the predictions of the IST, as per the calculations in Section~\ref{sec:scattering_problem}.

An even more dramatic instance of the same phenomenon arises in the case of  $\alpha=\pi$, as depicted in the inset of Fig.~\ref{fig:h_05_a_pi_th_0}(a).
 Indeed, the  spatiotemporal evolution of  this configuration captures the formation of
two counter-propagating dispersive shock waves whose downstream soliton emission~\cite{Dutton2001,Katsimiga2018} is illustrated in Fig.~\ref{fig:h_05_a_pi_th_0}(a).
 As in the case shown in Fig.~\ref{fig:h_05_a_0_th_0}(a), these shock waves  interact with the newly formed dark solitons, an interaction that  is most pronounced for the  the two central nearly black  solitons visible  in the inset of Fig.~\ref{fig:h_05_a_pi_th_0}(a).
For both cases, close inspection of the relevant insets indeed reveals that solitons with
positive velocities are initially formed at $x \approx -5$.
 On the other hand, the negative velocity ones arise symmetrically at
$x \approx 5$. 
Despite the much more involved soliton generation, in both cases our simulations almost perfectly match the analytical predictions when we set the origin of the latter at $x=0$ [see the identified zeros in Fig.~\ref{fig:h_05_a_0_th_0}(b) and Fig.~\ref{fig:h_05_a_pi_th_0}(b), respectively].
Our results  continue to hold  even for significantly larger evolution times than those depicted herein.
It is also at these later times, and in particular around $t\approx 1000$, that the two central dark solitons, whose zeros are identified at $k_1 = \pm 0.0045$ [see the $k_1>0$ in Fig.~\ref{fig:h_05_a_pi_th_0}(b)],
begin to repel~\cite{Theocharis2010}  one another  effectively, given their opposite but extremely small in magnitude velocities (results not shown here for brevity).
Finally, due to the above-described dynamics, both the instantaneous velocities, $v$ [Fig.~\ref{fig:h_05_a_0_th_0}(c), Fig.~\ref{fig:h_05_a_pi_th_0}(c)], and the amplitudes, $A_d$ [Fig.~\ref{fig:h_05_a_0_th_0}(d), Fig.~\ref{fig:h_05_a_pi_th_0}(d)], of all three pairs of solitons formed in both scenarios acquire their expected nearly constant trend for $t\geq 25$.

\begin{figure}[t]
    \centering
    \includegraphics[width=\columnwidth]{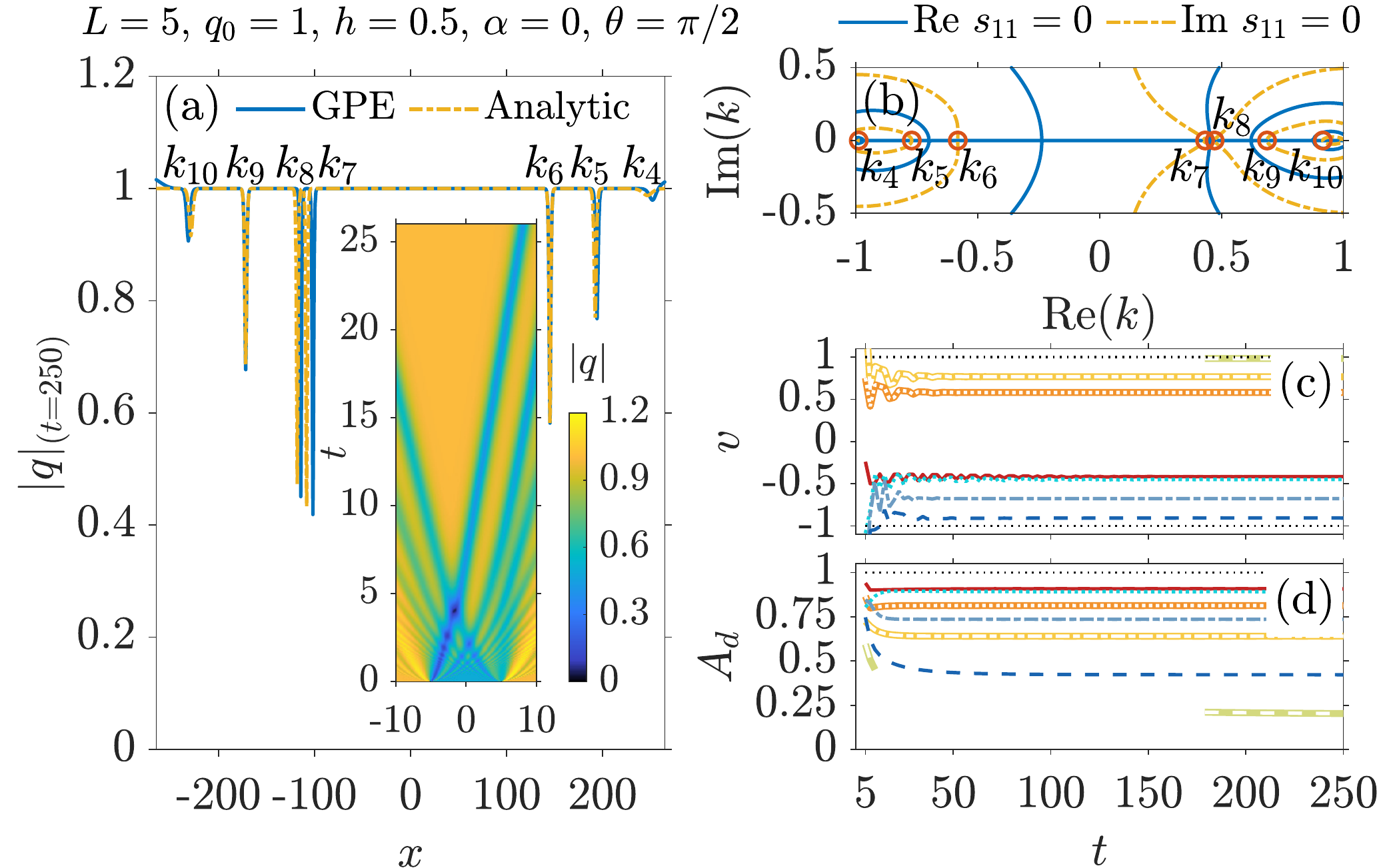}
    \caption{Same as Fig.~\ref{fig:h_0_a_0_th_0} but for $L=5$, $q_o=1$, $\theta=\pi/2$, $h=0.5$ and $\alpha=0$
    [see the legends in Fig.~\ref{fig:roots_Lh_q_1_a_0_th_pi2}].
    In this case (b) depicts all the complex $k$--plane.
    From left to right the zeros, $k_o$, in (b) that lead to the solitons formed in (a) are located at:
    $k_4=-0.9750$, $k_5=-0.7700$, $k_6=-0.5814$, $k_7=0.4329$, $k_8=0.4718$, $k_9=0.6869$, $k_{10}=0.9155$.}
    \label{fig:h_05_a_0_th_pi2}
\end{figure}

\smallskip
\textbf{ Other configurations.}
In all cases discussed so far, the initial configuration gave rise to a symmetric distribution of solitons.
We now explore a scenario corresponding  to an OP initial configuration, the analytical predictions of which can be found in Fig.~\ref{fig:roots_Lh_q_1_a_0_th_pi2}(e).
 The corresponding dynamical process is  illustrated in Figs.~\ref{fig:h_05_a_0_th_pi2}(a)--(d).
In contrast to the previously discussed IP box-type configurations, in the present case, since both $\Delta\theta=\pi$ and $\Delta\theta_\pm=\pi/2$, we do expect an asymmetric distribution of the zeros, $k_o$, and thus asymmetrically produced dark solitons.
Both expectations are confirmed and shown in Figs.~\ref{fig:h_05_a_0_th_pi2}(a) and \ref{fig:h_05_a_0_th_pi2}(b).
In particular, seven distinct solitons are nucleated in Fig.~\ref{fig:h_05_a_0_th_pi2}(a), with each of them corresponding to each of the seven distinct solutions shown in Fig.~\ref{fig:h_05_a_0_th_pi2}(b).
The  spatiotemporal evolution of $|q|$ at early times [inset of Fig.~\ref{fig:h_05_a_0_th_pi2}(a)]
shows that  three of them have $v=-k_o>0$ and four $v=-k_o<0$.
This  asymmetric generation of matter-waves entails also the largest deviations between our analytical findings and the numerically obtained ones.
This can be easily inferred by inspecting either the profiles or even better the estimated velocities, $v$, and amplitudes, $A_d$, of the ensuing waves illustrated respectively in Figs.~\ref{fig:h_05_a_0_th_pi2}(c) and \ref{fig:h_05_a_0_th_pi2}(d).
For instance, the fastest soliton, $k_4=-0.9750$, bears  such a  small amplitude that renders it indistinguishable from the background radiation
for times up to $t\approx 180$.
As such, the  corresponding  $v$ and $A_d$ are not depicted in Figs.~\ref{fig:h_05_a_0_th_pi2}(c) and \ref{fig:h_05_a_0_th_pi2}(d), respectively, until $t>180$.
Yet, another example refers to the solitons labelled $k_7$ and $k_8$. Namely, the two entities that are tightly close to one another [see here Eq.~\eqref{eq:s11_h_q_L_0} and Eq~\eqref{eq:s11_h_inf}]
and thus interact continuously with each other.
It is this continuous interaction that holds for $t\gtrsim 1000$, before the soliton repulsion sets in, to which the discrepancy in the amplitudes observed at $t=250$
is attributed [Fig.~\ref{fig:h_05_a_0_th_pi2}(a)].
Even though the largest deviation between our analytical predictions, provided by the zeros of Eq.~\eqref{eq:s11}, and our numerical findings is found for the aforementioned
asymmetric initial  configurations,  it still lies within our numerical precision, i.e. $\delta = \pm 0.01$.

We  also explored cases for which $h$ is close to $q_o$ but $h \geq q_o$.
Here, our results are found to be consistent with the limiting cases discussed in Sec.~\ref{sec:analogies}.
In particular, $\theta=\alpha=0$ leads to sound wave emission but no soliton production.
For $\theta=0$ and $\alpha=\pi$ the creation of two almost black solitons ($k_o\approx 0$)
located at $x=\pm (L+\epsilon)$, where $\epsilon>0$ is a small displacement caused by the emission of radiation, is seen.
Last, $\theta=\pi/2$ and $\alpha=0$ ($\theta=\pi/2$ and $\alpha=\pi$) results into two nearly equal zeros, with $\Re k_o > 0$ ($\Re k_o<0$).


\subsection{Nucleation of dark soliton trains:  With confinement} 
\label{sec:sol_trap}

We now aim  to generalize our findings  by taking into account  the presence of a harmonic confinement
that is naturally introduced in BEC experiments~\cite{Becker2008,Katsimiga2020,Bersano2018}.
To this end, for the numerical considerations to be presented below we turn on the harmonic potential introduced in Eq.~(\ref{eq:GPE}) and we further fix the trapping frequency to $\Omega=0.01$~\cite{Katsimiga2020}.
The latter choice, besides its experimental relevance, is also an optimal one since
it allows for properly handling the sound wave emission that takes place at the initial stages
of the interference process.
Indeed, for tighter trappings the radiation emitted remains also trapped and, as such, multiple collisions of the generated dark solitons with these sound waves would result in a much more involved dynamical evolution of the nucleated matter waves.
Yet, another important point worth mentioning here refers to the analytical estimates
regarding the soliton generation provided by solving the direct scattering problem (see Sec.~\ref{sec:scattering_problem}).
Specifically, in the trap setting under consideration these estimates can serve as
approximate ones, since for instance the NZBC, which in turn define the asymptotic behavior of the solitons formed in terms of amplitude, velocity and location, cannot be fulfilled.
However, as we shall show later on, the strength of the analytical predictions is not limited to the
homogeneous setup but provides a particularly insightful tool for the confined case as well.

\begin{figure}[t]
\centering
\includegraphics[width=\columnwidth]{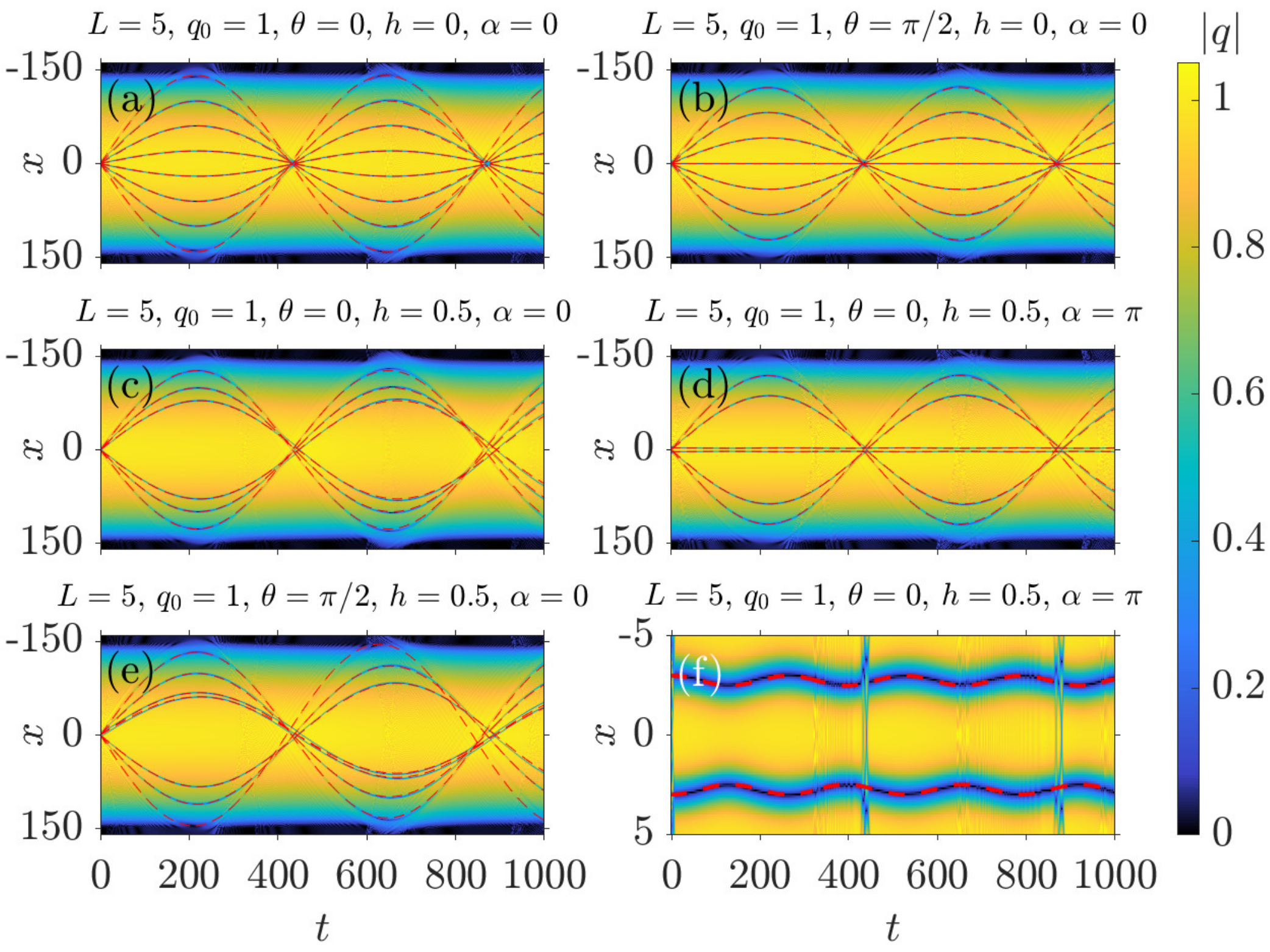}
\caption{(a)--(f) Spatiotemporal evolution of $|q|$ for distinct choices of
the involved parameters $L$, $q_o$, $h$, $\alpha$ and $\theta$ (see legends).
Dashed lines correspond to the fitted trajectories upon monitoring during evolution the center of mass
of each dark soliton (see text).
(f) Magnified version of (d) which captures the two dark solitons that are symmetrically placed around the trap center ($x=0$).
The other four solitons that appear at the collision points are stretched due to the zoom.
The trapping frequency is fixed to $\Omega=0.01$.}
\label{fig:trap}
\end{figure}

\begin{figure}[t]
\centering
\includegraphics[width=\columnwidth]{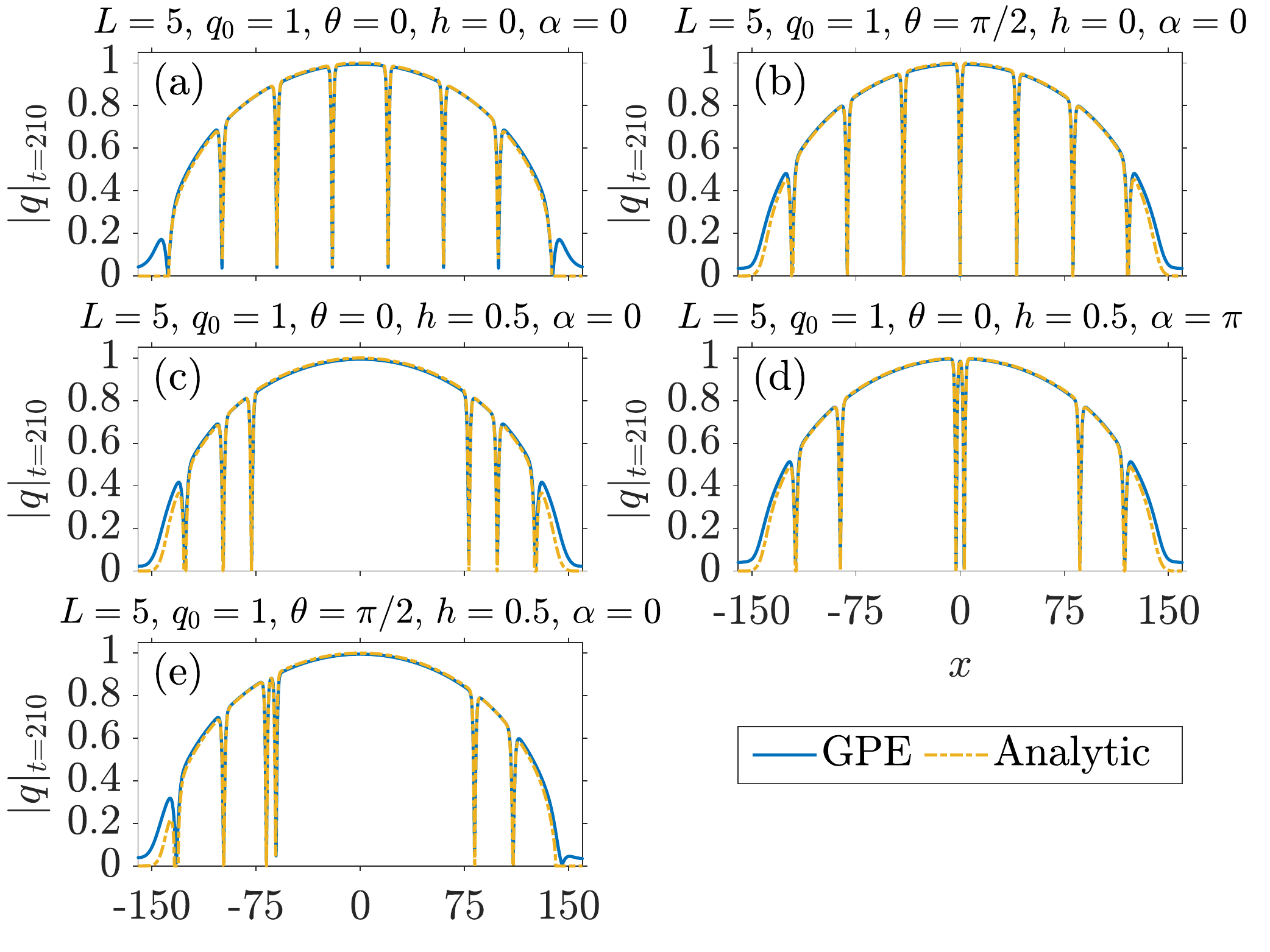}
\caption{(a)--(f) Profile snapshots of $|q|$ at $t=210$ for distinct choices of
the involved parameters $L$, $q_o$, $h$, $\alpha$ and $\theta$ (see legends).
The snapshots from (a) to (e) correspond to the relevant in each case dynamics presented in Figs.~\ref{fig:trap}(a)--(e) respectively.
The trapping frequency is fixed to $\Omega=0.01$.}
\label{fig:trap_profs}
\end{figure}

In the present setting, in order to induce the dynamics we initially find, by using imaginary time propagation, the ground state of the scalar system [Eq.~(\ref{eq:GPE})].
We then embed  in it  the wavefunction of Eq.(\ref{eq:initial_conditions}).
A schematic illustration of the aforementioned initial state is illustrated in Fig.~\ref{fig:IC}(b).
Moreover, in order to offer a direct comparison between the homogeneous and the confined cases,
we consider as representative examples five distinct selections of the involved parameters.
Namely, $L=5$, $q_o=1$ while $h=\qty{0, 0.5}$, $\alpha=\qty{0, \pi}$ and $\theta=\qty{0, \pi/2}$
(see also the relevant discussion around Figs.~\ref{fig:h_0_a_0_th_0}--\ref{fig:h_05_a_0_th_pi2} in Sec.~\ref{sec:sol_GPE}).
Our results are summarized in Figs.~\ref{fig:trap}(a)--(f) and Figs.~\ref{fig:trap_profs}(a)--(e),
as well as in Table~\ref{tab:freqs}.

\begin{table*}[t!]
    \centering
    \begin{tabular}{c c c | c c c | c c c | c c c | c c c}

        \hline\hline
        \multicolumn{3}{c}{$h=0,\alpha=0,\theta=0$} &
        \multicolumn{3}{|c}{$h=0,\alpha=0,\theta=\pi/2$} &
        \multicolumn{3}{|c}{$h=0.5,\alpha=0,\theta=0$} &
        \multicolumn{3}{|c}{$h=0.5,\alpha=\pi,\theta=0$} &
        \multicolumn{3}{|c}{$h=0.5,\alpha=0,\theta=\pi/2$} \\
        \hline\hline
        $k_o$ & $\omega_\textrm{num}$ & $\varepsilon_o$ &
        $k_o$ & $\omega_\textrm{num}$ & $\varepsilon_o$ &
        $k_o$ & $\omega_\textrm{num}$ & $\varepsilon_o$ &
        $k_o$ & $\omega_\textrm{num}$ & $\varepsilon_o$ &
        $k_o$ & $\omega_\textrm{num}$ & $\varepsilon_o$ \\
        \hline
        $k_1$ & 0.007184 & 0.016 &
        $k_7$ & 0 & -- &
        $k_1$ & 0.007043 & 0.004 &
        $k_1^{\dagger}$  & 0.023905 & 0.023 &
        $k_4$ & 0.007330 & 0.036
        \\
        $k_2$ & 0.007229 & 0.022 &
        $k_8$ & 0.007228 & 0.022 &
        $k_2$ & 0.007169 & 0.014 &
        $k_2$ & 0.007145 & 0.011 &
        $k_5$ & 0.007226 & 0.022
        \\
        $k_3$ & 0.007267 & 0.028 &
        $k_9$ & 0.007233 & 0.023 &
        $k_3$ & 0.007265 & 0.027 &
        $k_3$ & 0.007253 & 0.026 &
        $k_6$ & 0.007063 & 0.001
        \\
        $k_4$ & 0.007330 & 0.036 &
        $k_{10}$ & 0.007265 & 0.027 &
        & & &
        & & &
        $k_7$ & 0.007077 & 0.0008
        \\
        & & &
        & & &
        & & &
        & & &
        $k_8$ & 0.007077 & 0.0008
        \\
        & & &
        & & &
        & & &
        & & &
        $k_9$ & 0.007246 & 0.025
        \\
        & & &
        & & &
        & & &
        & & &
        $k_{10}$ & 0.007297 & 0.032
        \\
        \hline
    \end{tabular}
    \caption{Numerically obtained oscillation frequencies $\omega_\textrm{num}$.
    Each column contains the  parameter characterizing each soliton ($k_o$) and the  relative error ($\varepsilon_o$) with respect to the analytical
    prediction for the frequency of oscillation of a single dark soliton, i.e. $\omega_o=\Omega/\sqrt{2}$~\cite{Frantzeskakis2010}
    for different variations of the system's parameters.
    From left to right each column corresponds to Figs.~\ref{fig:trap}(a) up to Figs.~\ref{fig:trap}(e), respectively.
    Other parameters used are $L=5$, $q_o=1$ and $\Omega=0.01$.
    $^\dagger$See Eq.~\eqref{eq:freq_couple} and the discussion around.
    }
    \label{tab:freqs}
\end{table*}

In particular, Figs.~\ref{fig:trap}(a)--(e) illustrate the spatiotemporal evolution of $|q|$,
for different parametric variations. Additionally, Figs.~\ref{fig:trap_profs}(a)--(e) are the corresponding profile snapshots of $|q|$ at $t=210$ for each selection of parameters,  together with the relevant analytical estimates.
In general, it is found that the number of the dark solitons formed in each dynamical process
is the same as in the homogeneous case, but most importantly this formation is adequately described
by the analytical predictions [see Eq.~\eqref{eq:num_solutions}].
For instance, Figs.~\ref{fig:trap}(a) and Figs.~\ref{fig:trap_profs}(a)
show the generation of four pairs of dark solitons which is exactly the number of matter waves
that are predicted and observed for the homogeneous counterpart of this parameter selection illustrated in Fig.~\ref{fig:h_0_a_0_th_0}. Notice here the very good agreement
between the analytical predictions and the dynamically formed dark solitons.
This outcome holds equally also for the dynamical processes shown in Figs.~\ref{fig:trap}(b)--(d) and
Figs.~\ref{fig:trap_profs}(b)--(d) [cf. Figs.~\ref{fig:h_0_a_0_th_pi2}--\ref{fig:h_05_a_pi_th_0}, respectively].
Here, according to our homogeneous findings, three pairs of dark solitons are expected and indeed nucleate symmetrically around the trap center.
Notice also the central black soliton in the former of these processes.
Only one difference is worth commenting on, namely the last of the aforementioned cases [Fig.~\ref{fig:trap}(d)].
By monitoring the dynamical evolution of the pair of dark solitons that are closer to the trap center,
a magnified version of which is provided in Fig.~\ref{fig:trap}(f), it is found that these two dark solitons instead of executing large amplitude oscillations as the remaining pairs do, they lock into an out-of-phase oscillation mode, similar to the ones explored previously (including experimentally) in the works of, e.g.,~\cite{Weller2008,Theocharis2010}. 
As our last case example, in Fig.~\ref{fig:trap}(e) we show the dynamical evolution of the system for
parameters that lead to asymmetric soliton generation analogous to the one found in the homogeneous scenario [see Fig.~\ref{fig:h_05_a_0_th_pi2}].
Also in this case the number of dark solitons coincides with the one found in the homogeneous setting, with seven such entities being generated.
Even more importantly here, it is not only the number of nucleated states that is in accordance with the analytical predictions discussed in the homogeneous case, but also the relative position of the evolved states.
The latter is almost perfectly captured by the analytical solutions, but here during the oscillation of the asymmetric solitons formed [compare the panels of Fig.~\ref{fig:trap_profs}(e) to Fig.~\ref{fig:h_05_a_0_th_pi2}].
Note also that while the solitons corresponding to the solutions $k_7$ and $k_8$ shown in Figs.~\ref{fig:h_05_a_0_th_pi2}(a)--(b) propagate parallel to each other but eventually, due to repulsion, they will  separate out,  this is not the case for their trapped analogues shown in Fig.~\ref{fig:trap}(e) and Fig.~\ref{fig:trap_profs}(e), which oscillate  nearly parallel to each other, given the effect of the confining potential. 

In order to further shed light on the observed in-trap dynamics of the  dark solitons
generated in each case, we once more follow the center of mass, $x_{CM}$, of each entity for evolution times up to $t=3000$.
The numerically obtained oscillation frequencies, $\omega_\textrm{num}$, are included in Table~\ref{tab:freqs}.
In particular, Table~\ref{tab:freqs}, contains $\omega_\textrm{num}$ for each soliton that can in turn be compared to the (asymptotic) analytical prediction $\omega_o\equiv\Omega/\sqrt{2}=0.007071$  within
the so-called Thomas-Fermi regime where $q_o \gg \Omega$~\cite{Busch2000,Frantzeskakis2010}.
From left to right, each column of Table~\ref{tab:freqs} corresponds to Figs.~\ref{fig:trap}(a)--(e), respectively.
Additionally, the different solutions are denoted by the different zeros, $k_o$, identified by the scattering problem [see the notation introduced in Figs.~\ref{fig:h_0_a_0_th_0}--\ref{fig:h_05_a_0_th_pi2}].
Evidently, the faster moving solitons ($k_o\approx c$), see e.g. the outermost illustrated in
Fig.~\ref{fig:trap}(a) corresponding to the solution labelled $k_4$ in the first column of Table~\ref{tab:freqs}, have the largest $\omega_\textrm{num}$ and also the maximum deviation, $\varepsilon_o=|\omega_\textrm{num}-\omega_o|/\omega_o$, from the analytical prediction.
In some cases, such waves are indistinguishable from the radiation itself.
For these cases, we were not able to trace the center of mass of the ensuing soliton and thus obtain its oscillation frequency.
One such example corresponds to the fastest soliton shown in Fig.~\ref{fig:trap}(e), whose solution $k_4$ is depicted in the fifth column of Table~\ref{tab:freqs}, for which we determined $\omega_\textrm{num}$ manually.
It turns out that in all cases investigated herein, the maximum discrepancy between $\omega_\textrm{num}$ and $\omega_o$ is $\varepsilon_o=3.6\%$
(see $k_4$ in the first and fifth columns), while the minimum is $\varepsilon_o=0.08\%$ (see $k_7$ and $k_8$ in the fifth column).
Recalling now that $\omega_o$ is the oscillation frequency of a single dark soliton within the parabolic trap when slightly displaced from its equilibrium position, the observed discrepancies
can be attributed to: (i) the existence of more than one dark  solitons; 
(ii) the interaction of the dark solitons with the sound waves emitted during the dynamics;
(ii) The interactions among one another.  These effects have been studied previously in some of the above cited works, such as~\cite{Weller2008,Theocharis2010} and hence are not examined further here.
However, we can use their previous results to very accurately describe the out-of-phase oscillations from the soliton pair shown in Fig.~\ref{fig:trap}(f), for which we numerically obtained an oscillation frequency $\omega_\textrm{num}=0.023905$.
From~\cite{Theocharis2010}, the oscillation frequency of two solitons performing small out-of-phase oscillations around their equilibrium positions reads
\begin{align}
    \omega_{OP}^2=\omega_o^2+32q_o^2e^{-4q_o|x_\pm|}\,,
    \label{eq:freq_couple}
\end{align}
with the equilibrium positions, $x_\pm$, given by
\begin{align}
    x_\pm=\pm\frac{1}{4q_o}w\qty(\frac{32q_o^4}{\omega_o^2})\,,
    \label{eq:position_couple}
\end{align}
where $w(z)$ is the Lambert's $w$ function defined as the inverse of $z(w)=we^w$.
Then, Eq.~\eqref{eq:freq_couple} yields $\omega_{OP}=0.024487$.
This result is in very good agreement with the numerically found frequency, which presents only a relative error $\varepsilon_o=0.023$.


\section{Conclusions and Perspectives} \label{sec:conclusions}

In this work,  we have investigated  the  on-demand  nucleation of dark soliton trains arising in a 1D repulsively interacting scalar BEC system both in the absence and in the presence of a harmonic trap.
In particular, by utilizing box-shaped initial  configurations,  we have shown that it is possible
to a-priori predict not only the number of nucleated dark matter-waves, but also their
amplitudes, velocities and positions.
We have done so by initially considering the integrable version of the problem, namely the defocusing NLS equation.
For this model and for the aforementioned flexible initial wavefunction the direct scattering
problem has been solved analytically.
The direct relation of the discrete eigenvalues of the latter with the velocities and amplitudes of the emergent dark solitons has been showcased, while the exact soliton solutions are systematically extracted via IST.

By considering a wide range of parametric selections we have shown that the number and the symmetric or asymmetric  distribution of the nucleated soliton trains can be tailored upon suitable adjustment of the initial  configuration  parameters.
In general, and also in line with earlier predictions based on interference processes~\cite{Romero-Ros2019}, it is found that wider box-type configurations result in larger soliton trains.
However, narrower box-type configurations, resembling, in turn, phase imprint techniques that create defects within a BEC~\cite{Dutton2001}, lead to smaller soliton trains.
We have explored different types of configurations involving shallow boxes, as well as two entirely separated condensates. 
Also, asymmetrically distributed dark trains can be dynamically realized when considering e.g. shallow OP initial  configurations. 
Here, slowly-interacting  dark solitons coexisting with slow and extremely fast ones arise.  
In all the cases considered for the integrable defocusing NLS without a trap, our analytical findings are supported by the direct dynamical evolution of the scalar system. 
In particular, the velocities and amplitudes of the emergent soliton trains are traced during evolution  and both  approach the analytical predictions asymptotically, highlighting an excellent agreement between the two.
Finally we also appreciated the strength of our analytical predictions
even in the presence of a harmonic trap.
Our findings for all cases investigated in the latter setting closely followed the
ones identified in the homogeneous setup, with the anticipated modifications,  in each case scenario, in the amplitudes and velocities of the emitted dark soliton trains due to the presence of the trap.
Remarkable agreement between the analytical estimates and our numerical findings is exposed, with deviations regarding e.g. the estimated oscillation frequency of each nucleated matter-wave being less than $4\%$.

An immediate extension of this work points towards a  richer  system, consisting of two- component~\cite{Becker2008,Katsimiga2018a} or even three-component BECs~\cite{Bersano2018,Romero-Ros2019}.
In this regard, while recent works already considered multi-component BEC setups with box-type
initial  configurations~\cite{Romero-Ros2019}, revealing, among  other things,  the generation of dark-bright solitons trains, a systematic analytical treatment of the problem is still lacking.
Yet, another interesting perspective would be to generalize the diagnostics utilized herein in
higher dimensions.  There, naturally the toolbox of integrability is no longer available. Nevertheless, in this setting, topological excitations may be expected to emerge as a result of the interference process, in the presence of suitable phase structure, as has been shown, e.g.,
in the experiments of~\cite{Scherer2007}. 

\section{Acknowledgements}
P.G.K. is grateful to the Leverhulme Trust and to the Alexander von Humboldt Foundation for support
and to the Mathematical Institute of the University of Oxford for its hospitality.
A.R.-R and P.S. gratefully acknowledge financial support by the Deutsche
Forschungsgemeinschaft (DFG) in the framework of the SFB 925 ``Light induced dynamics and control of
correlated quantum systems".

\bibliographystyle{apsrev4-1}
\bibliography{biblio.bib}

\begin{thebibliography}{69}%
\makeatletter
\providecommand \@ifxundefined [1]{%
 \@ifx{#1\undefined}
}%
\providecommand \@ifnum [1]{%
 \ifnum #1\expandafter \@firstoftwo
 \else \expandafter \@secondoftwo
 \fi
}%
\providecommand \@ifx [1]{%
 \ifx #1\expandafter \@firstoftwo
 \else \expandafter \@secondoftwo
 \fi
}%
\providecommand \natexlab [1]{#1}%
\providecommand \enquote  [1]{``#1''}%
\providecommand \bibnamefont  [1]{#1}%
\providecommand \bibfnamefont [1]{#1}%
\providecommand \citenamefont [1]{#1}%
\providecommand \href@noop [0]{\@secondoftwo}%
\providecommand \href [0]{\begingroup \@sanitize@url \@href}%
\providecommand \@href[1]{\@@startlink{#1}\@@href}%
\providecommand \@@href[1]{\endgroup#1\@@endlink}%
\providecommand \@sanitize@url [0]{\catcode `\\12\catcode `\$12\catcode
  `\&12\catcode `\#12\catcode `\^12\catcode `\_12\catcode `\%12\relax}%
\providecommand \@@startlink[1]{}%
\providecommand \@@endlink[0]{}%
\providecommand \url  [0]{\begingroup\@sanitize@url \@url }%
\providecommand \@url [1]{\endgroup\@href {#1}{\urlprefix }}%
\providecommand \urlprefix  [0]{URL }%
\providecommand \Eprint [0]{\href }%
\providecommand \doibase [0]{http://dx.doi.org/}%
\providecommand \selectlanguage [0]{\@gobble}%
\providecommand \bibinfo  [0]{\@secondoftwo}%
\providecommand \bibfield  [0]{\@secondoftwo}%
\providecommand \translation [1]{[#1]}%
\providecommand \BibitemOpen [0]{}%
\providecommand \bibitemStop [0]{}%
\providecommand \bibitemNoStop [0]{.\EOS\space}%
\providecommand \EOS [0]{\spacefactor3000\relax}%
\providecommand \BibitemShut  [1]{\csname bibitem#1\endcsname}%
\let\auto@bib@innerbib\@empty
\bibitem [{\citenamefont {Chabchoub}\ \emph {et~al.}(2013)\citenamefont
  {Chabchoub}, \citenamefont {Kimmoun}, \citenamefont {Branger}, \citenamefont
  {Hoffmann}, \citenamefont {Proment}, \citenamefont {Onorato},\ and\
  \citenamefont {Akhmediev}}]{Chabchoub2013}%
  \BibitemOpen
  \bibfield  {author} {\bibinfo {author} {\bibfnamefont {A.}~\bibnamefont
  {Chabchoub}}, \bibinfo {author} {\bibfnamefont {O.}~\bibnamefont {Kimmoun}},
  \bibinfo {author} {\bibfnamefont {H.}~\bibnamefont {Branger}}, \bibinfo
  {author} {\bibfnamefont {N.}~\bibnamefont {Hoffmann}}, \bibinfo {author}
  {\bibfnamefont {D.}~\bibnamefont {Proment}}, \bibinfo {author} {\bibfnamefont
  {M.}~\bibnamefont {Onorato}}, \ and\ \bibinfo {author} {\bibfnamefont
  {N.}~\bibnamefont {Akhmediev}},\ }\href {\doibase
  10.1103/PhysRevLett.110.124101} {\bibfield  {journal} {\bibinfo  {journal}
  {Phys. Rev. Lett.}\ }\textbf {\bibinfo {volume} {110}},\ \bibinfo {pages}
  {124101} (\bibinfo {year} {2013})}\BibitemShut {NoStop}%
\bibitem [{\citenamefont {Tong}\ \emph {et~al.}(2010)\citenamefont {Tong},
  \citenamefont {Wu}, \citenamefont {Carr},\ and\ \citenamefont
  {Kalinikos}}]{Tong2010}%
  \BibitemOpen
  \bibfield  {author} {\bibinfo {author} {\bibfnamefont {W.}~\bibnamefont
  {Tong}}, \bibinfo {author} {\bibfnamefont {M.}~\bibnamefont {Wu}}, \bibinfo
  {author} {\bibfnamefont {L.~D.}\ \bibnamefont {Carr}}, \ and\ \bibinfo
  {author} {\bibfnamefont {B.~A.}\ \bibnamefont {Kalinikos}},\ }\href {\doibase
  10.1103/PhysRevLett.104.037207} {\bibfield  {journal} {\bibinfo  {journal}
  {Phys. Rev. Lett.}\ }\textbf {\bibinfo {volume} {104}},\ \bibinfo {pages}
  {037207} (\bibinfo {year} {2010})}\BibitemShut {NoStop}%
\bibitem [{\citenamefont {Zakharov}\ and\ \citenamefont
  {Shabat}(1973)}]{Zakharov1973}%
  \BibitemOpen
  \bibfield  {author} {\bibinfo {author} {\bibfnamefont {V.~E.}\ \bibnamefont
  {Zakharov}}\ and\ \bibinfo {author} {\bibfnamefont {A.~B.}\ \bibnamefont
  {Shabat}},\ }\href@noop {} {\bibfield  {journal} {\bibinfo  {journal} {Sov.
  J. Exp. Theor. Phys.}\ }\textbf {\bibinfo {volume} {37}},\ \bibinfo {pages}
  {823} (\bibinfo {year} {1973})}\BibitemShut {NoStop}%
\bibitem [{\citenamefont {Corney}\ \emph {et~al.}(1997)\citenamefont {Corney},
  \citenamefont {Drummond},\ and\ \citenamefont {Liebman}}]{Corney1997}%
  \BibitemOpen
  \bibfield  {author} {\bibinfo {author} {\bibfnamefont {J.~F.}\ \bibnamefont
  {Corney}}, \bibinfo {author} {\bibfnamefont {P.~D.}\ \bibnamefont
  {Drummond}}, \ and\ \bibinfo {author} {\bibfnamefont {A.}~\bibnamefont
  {Liebman}},\ }\href {\doibase 10.1016/S0030-4018(97)00191-0} {\bibfield
  {journal} {\bibinfo  {journal} {Optics Communications}\ }\textbf {\bibinfo
  {volume} {140}},\ \bibinfo {pages} {211} (\bibinfo {year}
  {1997})}\BibitemShut {NoStop}%
\bibitem [{\citenamefont {Kivshar}\ and\ \citenamefont
  {{Luther-Davies}}(1998)}]{Kivshar1998}%
  \BibitemOpen
  \bibfield  {author} {\bibinfo {author} {\bibfnamefont {Y.~S.}\ \bibnamefont
  {Kivshar}}\ and\ \bibinfo {author} {\bibfnamefont {B.}~\bibnamefont
  {{Luther-Davies}}},\ }\href {\doibase 10.1016/S0370-1573(97)00073-2}
  {\bibfield  {journal} {\bibinfo  {journal} {Physics Reports}\ }\textbf
  {\bibinfo {volume} {298}},\ \bibinfo {pages} {81} (\bibinfo {year}
  {1998})}\BibitemShut {NoStop}%
\bibitem [{\citenamefont {Pethick}\ and\ \citenamefont
  {Smith}(2008)}]{Pethick2008}%
  \BibitemOpen
  \bibfield  {author} {\bibinfo {author} {\bibfnamefont {C.~J.}\ \bibnamefont
  {Pethick}}\ and\ \bibinfo {author} {\bibfnamefont {H.}~\bibnamefont
  {Smith}},\ }\href {\doibase 10.1017/CBO9780511802850} {\emph {\bibinfo
  {title} {Bose\textendash{{Einstein Condensation}} in {{Dilute Gases}}}}},\
  \bibinfo {edition} {2nd}\ ed.\ (\bibinfo  {publisher} {{Cambridge University
  Press}},\ \bibinfo {address} {{Cambridge}},\ \bibinfo {year}
  {2008})\BibitemShut {NoStop}%
\bibitem [{\citenamefont {Pitaevskii}\ and\ \citenamefont
  {Stringari}(2016)}]{Pitaevskii2016}%
  \BibitemOpen
  \bibfield  {author} {\bibinfo {author} {\bibfnamefont {L.}~\bibnamefont
  {Pitaevskii}}\ and\ \bibinfo {author} {\bibfnamefont {S.}~\bibnamefont
  {Stringari}},\ }\href@noop {} {\emph {\bibinfo {title} {Bose-{{Einstein
  Condensation}} and {{Superfluidity}}}}},\ Vol.\ \bibinfo {volume} {164}\
  (\bibinfo  {publisher} {{Oxford University Press}},\ \bibinfo {year}
  {2016})\BibitemShut {NoStop}%
\bibitem [{\citenamefont {Becker}\ \emph {et~al.}(2008)\citenamefont {Becker},
  \citenamefont {Stellmer}, \citenamefont {{Soltan-Panahi}}, \citenamefont
  {D{\"o}rscher}, \citenamefont {Baumert}, \citenamefont {Richter},
  \citenamefont {Kronj{\"a}ger}, \citenamefont {Bongs},\ and\ \citenamefont
  {Sengstock}}]{Becker2008}%
  \BibitemOpen
  \bibfield  {author} {\bibinfo {author} {\bibfnamefont {C.}~\bibnamefont
  {Becker}}, \bibinfo {author} {\bibfnamefont {S.}~\bibnamefont {Stellmer}},
  \bibinfo {author} {\bibfnamefont {P.}~\bibnamefont {{Soltan-Panahi}}},
  \bibinfo {author} {\bibfnamefont {S.}~\bibnamefont {D{\"o}rscher}}, \bibinfo
  {author} {\bibfnamefont {M.}~\bibnamefont {Baumert}}, \bibinfo {author}
  {\bibfnamefont {E.-M.}\ \bibnamefont {Richter}}, \bibinfo {author}
  {\bibfnamefont {J.}~\bibnamefont {Kronj{\"a}ger}}, \bibinfo {author}
  {\bibfnamefont {K.}~\bibnamefont {Bongs}}, \ and\ \bibinfo {author}
  {\bibfnamefont {K.}~\bibnamefont {Sengstock}},\ }\href {\doibase
  10.1038/nphys962} {\bibfield  {journal} {\bibinfo  {journal} {Nature
  Physics}\ }\textbf {\bibinfo {volume} {4}},\ \bibinfo {pages} {496} (\bibinfo
  {year} {2008})}\BibitemShut {NoStop}%
\bibitem [{\citenamefont {Frantzeskakis}(2010)}]{Frantzeskakis2010}%
  \BibitemOpen
  \bibfield  {author} {\bibinfo {author} {\bibfnamefont {D.~J.}\ \bibnamefont
  {Frantzeskakis}},\ }\href {\doibase 10.1088/1751-8113/43/21/213001}
  {\bibfield  {journal} {\bibinfo  {journal} {J. Phys. A: Math. Theor.}\
  }\textbf {\bibinfo {volume} {43}},\ \bibinfo {pages} {213001} (\bibinfo
  {year} {2010})}\BibitemShut {NoStop}%
\bibitem [{\citenamefont {Kevrekidis}\ \emph {et~al.}(2015)\citenamefont
  {Kevrekidis}, \citenamefont {Frantzeskakis},\ and\ \citenamefont
  {{Carretero-Gonz{\'a}lez}}}]{Kevrekidis2015}%
  \BibitemOpen
  \bibfield  {author} {\bibinfo {author} {\bibfnamefont {P.~G.}\ \bibnamefont
  {Kevrekidis}}, \bibinfo {author} {\bibfnamefont {D.~J.}\ \bibnamefont
  {Frantzeskakis}}, \ and\ \bibinfo {author} {\bibfnamefont {R.}~\bibnamefont
  {{Carretero-Gonz{\'a}lez}}},\ }\href {\doibase 10.1137/1.9781611973945}
  {\emph {\bibinfo {title} {The {{Defocusing Nonlinear Schr\"odinger
  Equation}}: From Dark Soliton to Vortices and Vortex Rings}}},\ Other
  {{Titles}} in {{Applied Mathematics}}\ (\bibinfo  {publisher} {{Society for
  Industrial and Applied Mathematics}},\ \bibinfo {year} {2015})\BibitemShut
  {NoStop}%
\bibitem [{\citenamefont {Kevrekidis}\ \emph {et~al.}(2007)\citenamefont
  {Kevrekidis}, \citenamefont {Frantzeskakis},\ and\ \citenamefont
  {{Carretero-Gonz{\'a}lez}}}]{Kevrekidis2007}%
  \BibitemOpen
  \bibfield  {author} {\bibinfo {author} {\bibfnamefont {P.~G.}\ \bibnamefont
  {Kevrekidis}}, \bibinfo {author} {\bibfnamefont {D.~J.}\ \bibnamefont
  {Frantzeskakis}}, \ and\ \bibinfo {author} {\bibfnamefont {R.}~\bibnamefont
  {{Carretero-Gonz{\'a}lez}}},\ }\href@noop {} {\emph {\bibinfo {title}
  {Emergent {{Nonlinear Phenomena}} in {{Bose}}-{{Einstein Condensates}}:
  {{Theory}} and {{Experiment}}}}},\ Vol.~\bibinfo {volume} {45}\ (\bibinfo
  {publisher} {{Springer Science \& Business Media}},\ \bibinfo {year}
  {2007})\BibitemShut {NoStop}%
\bibitem [{\citenamefont {Bloch}\ \emph {et~al.}(2008)\citenamefont {Bloch},
  \citenamefont {Dalibard},\ and\ \citenamefont {Zwerger}}]{Bloch2008}%
  \BibitemOpen
  \bibfield  {author} {\bibinfo {author} {\bibfnamefont {I.}~\bibnamefont
  {Bloch}}, \bibinfo {author} {\bibfnamefont {J.}~\bibnamefont {Dalibard}}, \
  and\ \bibinfo {author} {\bibfnamefont {W.}~\bibnamefont {Zwerger}},\ }\href
  {\doibase 10.1103/RevModPhys.80.885} {\bibfield  {journal} {\bibinfo
  {journal} {Rev. Mod. Phys.}\ }\textbf {\bibinfo {volume} {80}},\ \bibinfo
  {pages} {885} (\bibinfo {year} {2008})}\BibitemShut {NoStop}%
\bibitem [{\citenamefont {Huang}\ \emph {et~al.}(2001)\citenamefont {Huang},
  \citenamefont {Velarde},\ and\ \citenamefont {Makarov}}]{Huang2001}%
  \BibitemOpen
  \bibfield  {author} {\bibinfo {author} {\bibfnamefont {G.}~\bibnamefont
  {Huang}}, \bibinfo {author} {\bibfnamefont {M.~G.}\ \bibnamefont {Velarde}},
  \ and\ \bibinfo {author} {\bibfnamefont {V.~A.}\ \bibnamefont {Makarov}},\
  }\href {\doibase 10.1103/PhysRevA.64.013617} {\bibfield  {journal} {\bibinfo
  {journal} {Phys. Rev. A}\ }\textbf {\bibinfo {volume} {64}},\ \bibinfo
  {pages} {013617} (\bibinfo {year} {2001})}\BibitemShut {NoStop}%
\bibitem [{\citenamefont {Stellmer}\ \emph {et~al.}(2008)\citenamefont
  {Stellmer}, \citenamefont {Becker}, \citenamefont {{Soltan-Panahi}},
  \citenamefont {Richter}, \citenamefont {D{\"o}rscher}, \citenamefont
  {Baumert}, \citenamefont {Kronj{\"a}ger}, \citenamefont {Bongs},\ and\
  \citenamefont {Sengstock}}]{Stellmer2008}%
  \BibitemOpen
  \bibfield  {author} {\bibinfo {author} {\bibfnamefont {S.}~\bibnamefont
  {Stellmer}}, \bibinfo {author} {\bibfnamefont {C.}~\bibnamefont {Becker}},
  \bibinfo {author} {\bibfnamefont {P.}~\bibnamefont {{Soltan-Panahi}}},
  \bibinfo {author} {\bibfnamefont {E.-M.}\ \bibnamefont {Richter}}, \bibinfo
  {author} {\bibfnamefont {S.}~\bibnamefont {D{\"o}rscher}}, \bibinfo {author}
  {\bibfnamefont {M.}~\bibnamefont {Baumert}}, \bibinfo {author} {\bibfnamefont
  {J.}~\bibnamefont {Kronj{\"a}ger}}, \bibinfo {author} {\bibfnamefont
  {K.}~\bibnamefont {Bongs}}, \ and\ \bibinfo {author} {\bibfnamefont
  {K.}~\bibnamefont {Sengstock}},\ }\href {\doibase
  10.1103/PhysRevLett.101.120406} {\bibfield  {journal} {\bibinfo  {journal}
  {Phys. Rev. Lett.}\ }\textbf {\bibinfo {volume} {101}},\ \bibinfo {pages}
  {120406} (\bibinfo {year} {2008})}\BibitemShut {NoStop}%
\bibitem [{\citenamefont {Kamchatnov}\ and\ \citenamefont
  {Salerno}(2009)}]{Kamchatnov2009}%
  \BibitemOpen
  \bibfield  {author} {\bibinfo {author} {\bibfnamefont {A.~M.}\ \bibnamefont
  {Kamchatnov}}\ and\ \bibinfo {author} {\bibfnamefont {M.}~\bibnamefont
  {Salerno}},\ }\href {\doibase 10.1088/0953-4075/42/18/185303} {\bibfield
  {journal} {\bibinfo  {journal} {J. Phys. B: At. Mol. Opt. Phys.}\ }\textbf
  {\bibinfo {volume} {42}},\ \bibinfo {pages} {185303} (\bibinfo {year}
  {2009})}\BibitemShut {NoStop}%
\bibitem [{\citenamefont {Jezek}\ \emph {et~al.}(2016)\citenamefont {Jezek},
  \citenamefont {Capuzzi},\ and\ \citenamefont {Cataldo}}]{Jezek2016}%
  \BibitemOpen
  \bibfield  {author} {\bibinfo {author} {\bibfnamefont {D.~M.}\ \bibnamefont
  {Jezek}}, \bibinfo {author} {\bibfnamefont {P.}~\bibnamefont {Capuzzi}}, \
  and\ \bibinfo {author} {\bibfnamefont {H.~M.}\ \bibnamefont {Cataldo}},\
  }\href {\doibase 10.1103/PhysRevA.93.023601} {\bibfield  {journal} {\bibinfo
  {journal} {Phys. Rev. A}\ }\textbf {\bibinfo {volume} {93}},\ \bibinfo
  {pages} {023601} (\bibinfo {year} {2016})}\BibitemShut {NoStop}%
\bibitem [{\citenamefont {Hoefer}\ \emph {et~al.}(2011)\citenamefont {Hoefer},
  \citenamefont {Chang}, \citenamefont {Hamner},\ and\ \citenamefont
  {Engels}}]{Hoefer2011}%
  \BibitemOpen
  \bibfield  {author} {\bibinfo {author} {\bibfnamefont {M.~A.}\ \bibnamefont
  {Hoefer}}, \bibinfo {author} {\bibfnamefont {J.~J.}\ \bibnamefont {Chang}},
  \bibinfo {author} {\bibfnamefont {C.}~\bibnamefont {Hamner}}, \ and\ \bibinfo
  {author} {\bibfnamefont {P.}~\bibnamefont {Engels}},\ }\href {\doibase
  10.1103/PhysRevA.84.041605} {\bibfield  {journal} {\bibinfo  {journal} {Phys.
  Rev. A}\ }\textbf {\bibinfo {volume} {84}},\ \bibinfo {pages} {041605}
  (\bibinfo {year} {2011})}\BibitemShut {NoStop}%
\bibitem [{\citenamefont {Yan}\ \emph {et~al.}(2012)\citenamefont {Yan},
  \citenamefont {Chang}, \citenamefont {Hamner}, \citenamefont {Hoefer},
  \citenamefont {Kevrekidis}, \citenamefont {Engels}, \citenamefont
  {Achilleos}, \citenamefont {Frantzeskakis},\ and\ \citenamefont
  {Cuevas}}]{Yan2012}%
  \BibitemOpen
  \bibfield  {author} {\bibinfo {author} {\bibfnamefont {D.}~\bibnamefont
  {Yan}}, \bibinfo {author} {\bibfnamefont {J.~J.}\ \bibnamefont {Chang}},
  \bibinfo {author} {\bibfnamefont {C.}~\bibnamefont {Hamner}}, \bibinfo
  {author} {\bibfnamefont {M.}~\bibnamefont {Hoefer}}, \bibinfo {author}
  {\bibfnamefont {P.~G.}\ \bibnamefont {Kevrekidis}}, \bibinfo {author}
  {\bibfnamefont {P.}~\bibnamefont {Engels}}, \bibinfo {author} {\bibfnamefont
  {V.}~\bibnamefont {Achilleos}}, \bibinfo {author} {\bibfnamefont {D.~J.}\
  \bibnamefont {Frantzeskakis}}, \ and\ \bibinfo {author} {\bibfnamefont
  {J.}~\bibnamefont {Cuevas}},\ }\href {\doibase
  10.1088/0953-4075/45/11/115301} {\bibfield  {journal} {\bibinfo  {journal}
  {J. Phys. B: At. Mol. Opt. Phys.}\ }\textbf {\bibinfo {volume} {45}},\
  \bibinfo {pages} {115301} (\bibinfo {year} {2012})}\BibitemShut {NoStop}%
\bibitem [{\citenamefont {Bersano}\ \emph {et~al.}(2018)\citenamefont
  {Bersano}, \citenamefont {Gokhroo}, \citenamefont {Khamehchi}, \citenamefont
  {D'Ambroise}, \citenamefont {Frantzeskakis}, \citenamefont {Engels},\ and\
  \citenamefont {Kevrekidis}}]{Bersano2018}%
  \BibitemOpen
  \bibfield  {author} {\bibinfo {author} {\bibfnamefont {T.~M.}\ \bibnamefont
  {Bersano}}, \bibinfo {author} {\bibfnamefont {V.}~\bibnamefont {Gokhroo}},
  \bibinfo {author} {\bibfnamefont {M.~A.}\ \bibnamefont {Khamehchi}}, \bibinfo
  {author} {\bibfnamefont {J.}~\bibnamefont {D'Ambroise}}, \bibinfo {author}
  {\bibfnamefont {D.~J.}\ \bibnamefont {Frantzeskakis}}, \bibinfo {author}
  {\bibfnamefont {P.}~\bibnamefont {Engels}}, \ and\ \bibinfo {author}
  {\bibfnamefont {P.~G.}\ \bibnamefont {Kevrekidis}},\ }\href {\doibase
  10.1103/PhysRevLett.120.063202} {\bibfield  {journal} {\bibinfo  {journal}
  {Phys. Rev. Lett.}\ }\textbf {\bibinfo {volume} {120}},\ \bibinfo {pages}
  {063202} (\bibinfo {year} {2018})}\BibitemShut {NoStop}%
\bibitem [{\citenamefont {Denschlag}\ \emph {et~al.}(2000)\citenamefont
  {Denschlag}, \citenamefont {Simsarian}, \citenamefont {Feder}, \citenamefont
  {Clark}, \citenamefont {Collins}, \citenamefont {Cubizolles}, \citenamefont
  {Deng}, \citenamefont {Hagley}, \citenamefont {Helmerson}, \citenamefont
  {Reinhardt}, \citenamefont {Rolston}, \citenamefont {Schneider},\ and\
  \citenamefont {Phillips}}]{Denschlag2000}%
  \BibitemOpen
  \bibfield  {author} {\bibinfo {author} {\bibfnamefont {J.}~\bibnamefont
  {Denschlag}}, \bibinfo {author} {\bibfnamefont {J.~E.}\ \bibnamefont
  {Simsarian}}, \bibinfo {author} {\bibfnamefont {D.~L.}\ \bibnamefont
  {Feder}}, \bibinfo {author} {\bibfnamefont {C.~W.}\ \bibnamefont {Clark}},
  \bibinfo {author} {\bibfnamefont {L.~A.}\ \bibnamefont {Collins}}, \bibinfo
  {author} {\bibfnamefont {J.}~\bibnamefont {Cubizolles}}, \bibinfo {author}
  {\bibfnamefont {L.}~\bibnamefont {Deng}}, \bibinfo {author} {\bibfnamefont
  {E.~W.}\ \bibnamefont {Hagley}}, \bibinfo {author} {\bibfnamefont
  {K.}~\bibnamefont {Helmerson}}, \bibinfo {author} {\bibfnamefont {W.~P.}\
  \bibnamefont {Reinhardt}}, \bibinfo {author} {\bibfnamefont {S.~L.}\
  \bibnamefont {Rolston}}, \bibinfo {author} {\bibfnamefont {B.~I.}\
  \bibnamefont {Schneider}}, \ and\ \bibinfo {author} {\bibfnamefont {W.~D.}\
  \bibnamefont {Phillips}},\ }\href {\doibase 10.1126/science.287.5450.97}
  {\bibfield  {journal} {\bibinfo  {journal} {Science}\ }\textbf {\bibinfo
  {volume} {287}},\ \bibinfo {pages} {97} (\bibinfo {year} {2000})}\BibitemShut
  {NoStop}%
\bibitem [{\citenamefont {Anderson}\ \emph {et~al.}(2001)\citenamefont
  {Anderson}, \citenamefont {Haljan}, \citenamefont {Regal}, \citenamefont
  {Feder}, \citenamefont {Collins}, \citenamefont {Clark},\ and\ \citenamefont
  {Cornell}}]{Anderson2001}%
  \BibitemOpen
  \bibfield  {author} {\bibinfo {author} {\bibfnamefont {B.~P.}\ \bibnamefont
  {Anderson}}, \bibinfo {author} {\bibfnamefont {P.~C.}\ \bibnamefont
  {Haljan}}, \bibinfo {author} {\bibfnamefont {C.~A.}\ \bibnamefont {Regal}},
  \bibinfo {author} {\bibfnamefont {D.~L.}\ \bibnamefont {Feder}}, \bibinfo
  {author} {\bibfnamefont {L.~A.}\ \bibnamefont {Collins}}, \bibinfo {author}
  {\bibfnamefont {C.~W.}\ \bibnamefont {Clark}}, \ and\ \bibinfo {author}
  {\bibfnamefont {E.~A.}\ \bibnamefont {Cornell}},\ }\href {\doibase
  10.1103/PhysRevLett.86.2926} {\bibfield  {journal} {\bibinfo  {journal}
  {Phys. Rev. Lett.}\ }\textbf {\bibinfo {volume} {86}},\ \bibinfo {pages}
  {2926} (\bibinfo {year} {2001})}\BibitemShut {NoStop}%
\bibitem [{\citenamefont {Shomroni}\ \emph {et~al.}(2009)\citenamefont
  {Shomroni}, \citenamefont {Lahoud}, \citenamefont {Levy},\ and\ \citenamefont
  {Steinhauer}}]{Shomroni2009}%
  \BibitemOpen
  \bibfield  {author} {\bibinfo {author} {\bibfnamefont {I.}~\bibnamefont
  {Shomroni}}, \bibinfo {author} {\bibfnamefont {E.}~\bibnamefont {Lahoud}},
  \bibinfo {author} {\bibfnamefont {S.}~\bibnamefont {Levy}}, \ and\ \bibinfo
  {author} {\bibfnamefont {J.}~\bibnamefont {Steinhauer}},\ }\href {\doibase
  10.1038/nphys1177} {\bibfield  {journal} {\bibinfo  {journal} {Nat. Phys.}\
  }\textbf {\bibinfo {volume} {5}},\ \bibinfo {pages} {193} (\bibinfo {year}
  {2009})}\BibitemShut {NoStop}%
\bibitem [{\citenamefont {Burger}\ \emph {et~al.}(1999)\citenamefont {Burger},
  \citenamefont {Bongs}, \citenamefont {Dettmer}, \citenamefont {Ertmer},
  \citenamefont {Sengstock}, \citenamefont {Sanpera}, \citenamefont
  {Shlyapnikov},\ and\ \citenamefont {Lewenstein}}]{Burger1999}%
  \BibitemOpen
  \bibfield  {author} {\bibinfo {author} {\bibfnamefont {S.}~\bibnamefont
  {Burger}}, \bibinfo {author} {\bibfnamefont {K.}~\bibnamefont {Bongs}},
  \bibinfo {author} {\bibfnamefont {S.}~\bibnamefont {Dettmer}}, \bibinfo
  {author} {\bibfnamefont {W.}~\bibnamefont {Ertmer}}, \bibinfo {author}
  {\bibfnamefont {K.}~\bibnamefont {Sengstock}}, \bibinfo {author}
  {\bibfnamefont {A.}~\bibnamefont {Sanpera}}, \bibinfo {author} {\bibfnamefont
  {G.~V.}\ \bibnamefont {Shlyapnikov}}, \ and\ \bibinfo {author} {\bibfnamefont
  {M.}~\bibnamefont {Lewenstein}},\ }\href {\doibase
  10.1103/PhysRevLett.83.5198} {\bibfield  {journal} {\bibinfo  {journal}
  {Phys. Rev. Lett.}\ }\textbf {\bibinfo {volume} {83}},\ \bibinfo {pages}
  {5198} (\bibinfo {year} {1999})}\BibitemShut {NoStop}%
\bibitem [{\citenamefont {Dutton}\ \emph {et~al.}(2001)\citenamefont {Dutton},
  \citenamefont {Budde}, \citenamefont {Slowe},\ and\ \citenamefont
  {Hau}}]{Dutton2001}%
  \BibitemOpen
  \bibfield  {author} {\bibinfo {author} {\bibfnamefont {Z.}~\bibnamefont
  {Dutton}}, \bibinfo {author} {\bibfnamefont {M.}~\bibnamefont {Budde}},
  \bibinfo {author} {\bibfnamefont {C.}~\bibnamefont {Slowe}}, \ and\ \bibinfo
  {author} {\bibfnamefont {L.~V.}\ \bibnamefont {Hau}},\ }\href {\doibase
  10.1126/science.1062527} {\bibfield  {journal} {\bibinfo  {journal}
  {Science}\ }\textbf {\bibinfo {volume} {293}},\ \bibinfo {pages} {663}
  (\bibinfo {year} {2001})}\BibitemShut {NoStop}%
\bibitem [{\citenamefont {Engels}\ and\ \citenamefont
  {Atherton}(2007)}]{Engels2007}%
  \BibitemOpen
  \bibfield  {author} {\bibinfo {author} {\bibfnamefont {P.}~\bibnamefont
  {Engels}}\ and\ \bibinfo {author} {\bibfnamefont {C.}~\bibnamefont
  {Atherton}},\ }\href {\doibase 10.1103/PhysRevLett.99.160405} {\bibfield
  {journal} {\bibinfo  {journal} {Phys. Rev. Lett.}\ }\textbf {\bibinfo
  {volume} {99}},\ \bibinfo {pages} {160405} (\bibinfo {year}
  {2007})}\BibitemShut {NoStop}%
\bibitem [{\citenamefont {Reinhardt}\ and\ \citenamefont
  {Clark}(1997)}]{Reinhardt1997}%
  \BibitemOpen
  \bibfield  {author} {\bibinfo {author} {\bibfnamefont {W.~P.}\ \bibnamefont
  {Reinhardt}}\ and\ \bibinfo {author} {\bibfnamefont {C.~W.}\ \bibnamefont
  {Clark}},\ }\href {\doibase 10.1088/0953-4075/30/22/001} {\bibfield
  {journal} {\bibinfo  {journal} {J. Phys. B: At. Mol. Opt. Phys.}\ }\textbf
  {\bibinfo {volume} {30}},\ \bibinfo {pages} {785} (\bibinfo {year}
  {1997})}\BibitemShut {NoStop}%
\bibitem [{\citenamefont {Scott}\ \emph {et~al.}(1998)\citenamefont {Scott},
  \citenamefont {Ballagh},\ and\ \citenamefont {Burnett}}]{Scott1998}%
  \BibitemOpen
  \bibfield  {author} {\bibinfo {author} {\bibfnamefont {T.~F.}\ \bibnamefont
  {Scott}}, \bibinfo {author} {\bibfnamefont {R.~J.}\ \bibnamefont {Ballagh}},
  \ and\ \bibinfo {author} {\bibfnamefont {K.}~\bibnamefont {Burnett}},\ }\href
  {\doibase 10.1088/0953-4075/31/8/001} {\bibfield  {journal} {\bibinfo
  {journal} {J. Phys. B: At. Mol. Opt. Phys.}\ }\textbf {\bibinfo {volume}
  {31}},\ \bibinfo {pages} {329} (\bibinfo {year} {1998})}\BibitemShut
  {NoStop}%
\bibitem [{\citenamefont {Weller}\ \emph {et~al.}(2008)\citenamefont {Weller},
  \citenamefont {Ronzheimer}, \citenamefont {Gross}, \citenamefont {Esteve},
  \citenamefont {Oberthaler}, \citenamefont {Frantzeskakis}, \citenamefont
  {Theocharis},\ and\ \citenamefont {Kevrekidis}}]{Weller2008}%
  \BibitemOpen
  \bibfield  {author} {\bibinfo {author} {\bibfnamefont {A.}~\bibnamefont
  {Weller}}, \bibinfo {author} {\bibfnamefont {J.~P.}\ \bibnamefont
  {Ronzheimer}}, \bibinfo {author} {\bibfnamefont {C.}~\bibnamefont {Gross}},
  \bibinfo {author} {\bibfnamefont {J.}~\bibnamefont {Esteve}}, \bibinfo
  {author} {\bibfnamefont {M.~K.}\ \bibnamefont {Oberthaler}}, \bibinfo
  {author} {\bibfnamefont {D.~J.}\ \bibnamefont {Frantzeskakis}}, \bibinfo
  {author} {\bibfnamefont {G.}~\bibnamefont {Theocharis}}, \ and\ \bibinfo
  {author} {\bibfnamefont {P.~G.}\ \bibnamefont {Kevrekidis}},\ }\href
  {\doibase 10.1103/PhysRevLett.101.130401} {\bibfield  {journal} {\bibinfo
  {journal} {Phys. Rev. Lett.}\ }\textbf {\bibinfo {volume} {101}},\ \bibinfo
  {pages} {130401} (\bibinfo {year} {2008})}\BibitemShut {NoStop}%
\bibitem [{\citenamefont {Theocharis}\ \emph {et~al.}(2010)\citenamefont
  {Theocharis}, \citenamefont {Weller}, \citenamefont {Ronzheimer},
  \citenamefont {Gross}, \citenamefont {Oberthaler}, \citenamefont
  {Kevrekidis},\ and\ \citenamefont {Frantzeskakis}}]{Theocharis2010}%
  \BibitemOpen
  \bibfield  {author} {\bibinfo {author} {\bibfnamefont {G.}~\bibnamefont
  {Theocharis}}, \bibinfo {author} {\bibfnamefont {A.}~\bibnamefont {Weller}},
  \bibinfo {author} {\bibfnamefont {J.~P.}\ \bibnamefont {Ronzheimer}},
  \bibinfo {author} {\bibfnamefont {C.}~\bibnamefont {Gross}}, \bibinfo
  {author} {\bibfnamefont {M.~K.}\ \bibnamefont {Oberthaler}}, \bibinfo
  {author} {\bibfnamefont {P.~G.}\ \bibnamefont {Kevrekidis}}, \ and\ \bibinfo
  {author} {\bibfnamefont {D.~J.}\ \bibnamefont {Frantzeskakis}},\ }\href
  {\doibase 10.1103/PhysRevA.81.063604} {\bibfield  {journal} {\bibinfo
  {journal} {Phys. Rev. A}\ }\textbf {\bibinfo {volume} {81}},\ \bibinfo
  {pages} {063604} (\bibinfo {year} {2010})}\BibitemShut {NoStop}%
\bibitem [{\citenamefont {Brazhnyi}\ and\ \citenamefont
  {Kamchatnov}(2003)}]{Brazhnyi2003}%
  \BibitemOpen
  \bibfield  {author} {\bibinfo {author} {\bibfnamefont {V.~A.}\ \bibnamefont
  {Brazhnyi}}\ and\ \bibinfo {author} {\bibfnamefont {A.~M.}\ \bibnamefont
  {Kamchatnov}},\ }\href {\doibase 10.1103/PhysRevA.68.043614} {\bibfield
  {journal} {\bibinfo  {journal} {Phys. Rev. A}\ }\textbf {\bibinfo {volume}
  {68}},\ \bibinfo {pages} {043614} (\bibinfo {year} {2003})}\BibitemShut
  {NoStop}%
\bibitem [{\citenamefont {{Romero-Ros}}\ \emph {et~al.}(2019)\citenamefont
  {{Romero-Ros}}, \citenamefont {Katsimiga}, \citenamefont {Kevrekidis},\ and\
  \citenamefont {Schmelcher}}]{Romero-Ros2019}%
  \BibitemOpen
  \bibfield  {author} {\bibinfo {author} {\bibfnamefont {A.}~\bibnamefont
  {{Romero-Ros}}}, \bibinfo {author} {\bibfnamefont {G.~C.}\ \bibnamefont
  {Katsimiga}}, \bibinfo {author} {\bibfnamefont {P.~G.}\ \bibnamefont
  {Kevrekidis}}, \ and\ \bibinfo {author} {\bibfnamefont {P.}~\bibnamefont
  {Schmelcher}},\ }\href {\doibase 10.1103/PhysRevA.100.013626} {\bibfield
  {journal} {\bibinfo  {journal} {Phys. Rev. A}\ }\textbf {\bibinfo {volume}
  {100}},\ \bibinfo {pages} {013626} (\bibinfo {year} {2019})}\BibitemShut
  {NoStop}%
\bibitem [{\citenamefont {Demontis}\ \emph {et~al.}(2013)\citenamefont
  {Demontis}, \citenamefont {Prinari}, \citenamefont {van~der Mee},\ and\
  \citenamefont {Vitale}}]{Demontis2013}%
  \BibitemOpen
  \bibfield  {author} {\bibinfo {author} {\bibfnamefont {F.}~\bibnamefont
  {Demontis}}, \bibinfo {author} {\bibfnamefont {B.}~\bibnamefont {Prinari}},
  \bibinfo {author} {\bibfnamefont {C.}~\bibnamefont {van~der Mee}}, \ and\
  \bibinfo {author} {\bibfnamefont {F.}~\bibnamefont {Vitale}},\ }\href
  {\doibase 10.1111/j.1467-9590.2012.00572.x} {\bibfield  {journal} {\bibinfo
  {journal} {Stud. Appl. Math.}\ }\textbf {\bibinfo {volume} {131}},\ \bibinfo
  {pages} {1} (\bibinfo {year} {2013})}\BibitemShut {NoStop}%
\bibitem [{\citenamefont {Biondini}\ and\ \citenamefont
  {Prinari}(2014)}]{Biondini2014}%
  \BibitemOpen
  \bibfield  {author} {\bibinfo {author} {\bibfnamefont {G.}~\bibnamefont
  {Biondini}}\ and\ \bibinfo {author} {\bibfnamefont {B.}~\bibnamefont
  {Prinari}},\ }\href {\doibase 10.1111/sapm.12024} {\bibfield  {journal}
  {\bibinfo  {journal} {Stud. Appl. Math.}\ }\textbf {\bibinfo {volume}
  {132}},\ \bibinfo {pages} {138} (\bibinfo {year} {2014})}\BibitemShut
  {NoStop}%
\bibitem [{\citenamefont {Biondini}\ \emph {et~al.}(2016)\citenamefont
  {Biondini}, \citenamefont {Fagerstrom},\ and\ \citenamefont
  {Prinari}}]{Biondini2016a}%
  \BibitemOpen
  \bibfield  {author} {\bibinfo {author} {\bibfnamefont {G.}~\bibnamefont
  {Biondini}}, \bibinfo {author} {\bibfnamefont {E.}~\bibnamefont
  {Fagerstrom}}, \ and\ \bibinfo {author} {\bibfnamefont {B.}~\bibnamefont
  {Prinari}},\ }\href {\doibase 10.1016/j.physd.2016.04.003} {\bibfield
  {journal} {\bibinfo  {journal} {Physica D: Nonlinear Phenomena}\ }\textbf
  {\bibinfo {volume} {333}},\ \bibinfo {pages} {117} (\bibinfo {year}
  {2016})}\BibitemShut {NoStop}%
\bibitem [{\citenamefont {{Esp{\'i}nola-Rocha}}\ and\ \citenamefont
  {Kevrekidis}(2009)}]{Espinola-Rocha2009}%
  \BibitemOpen
  \bibfield  {author} {\bibinfo {author} {\bibfnamefont {J.~A.}\ \bibnamefont
  {{Esp{\'i}nola-Rocha}}}\ and\ \bibinfo {author} {\bibfnamefont
  {P.}~\bibnamefont {Kevrekidis}},\ }\href {\doibase
  10.1016/j.matcom.2009.08.022} {\bibfield  {journal} {\bibinfo  {journal}
  {Math. Comput. Simul.}\ }\textbf {\bibinfo {volume} {80}},\ \bibinfo {pages}
  {693} (\bibinfo {year} {2009})}\BibitemShut {NoStop}%
\bibitem [{\citenamefont {Gredeskul}\ and\ \citenamefont
  {Kivshar}(1989)}]{Gredeskul1989}%
  \BibitemOpen
  \bibfield  {author} {\bibinfo {author} {\bibfnamefont {S.~A.}\ \bibnamefont
  {Gredeskul}}\ and\ \bibinfo {author} {\bibfnamefont {Y.~S.}\ \bibnamefont
  {Kivshar}},\ }\href {\doibase 10.1103/PhysRevLett.62.977} {\bibfield
  {journal} {\bibinfo  {journal} {Phys. Rev. Lett.}\ }\textbf {\bibinfo
  {volume} {62}},\ \bibinfo {pages} {977} (\bibinfo {year} {1989})}\BibitemShut
  {NoStop}%
\bibitem [{\citenamefont {Swartzlander}\ \emph {et~al.}(1991)\citenamefont
  {Swartzlander}, \citenamefont {Andersen}, \citenamefont {Regan},
  \citenamefont {Yin},\ and\ \citenamefont {Kaplan}}]{Swartzlander1991}%
  \BibitemOpen
  \bibfield  {author} {\bibinfo {author} {\bibfnamefont {G.~A.}\ \bibnamefont
  {Swartzlander}}, \bibinfo {author} {\bibfnamefont {D.~R.}\ \bibnamefont
  {Andersen}}, \bibinfo {author} {\bibfnamefont {J.~J.}\ \bibnamefont {Regan}},
  \bibinfo {author} {\bibfnamefont {H.}~\bibnamefont {Yin}}, \ and\ \bibinfo
  {author} {\bibfnamefont {A.~E.}\ \bibnamefont {Kaplan}},\ }\href {\doibase
  10.1103/PhysRevLett.66.1583} {\bibfield  {journal} {\bibinfo  {journal}
  {Phys. Rev. Lett.}\ }\textbf {\bibinfo {volume} {66}},\ \bibinfo {pages}
  {1583} (\bibinfo {year} {1991})}\BibitemShut {NoStop}%
\bibitem [{\citenamefont {Ostrovskaya}\ \emph {et~al.}(1999)\citenamefont
  {Ostrovskaya}, \citenamefont {Kivshar}, \citenamefont {Chen},\ and\
  \citenamefont {Segev}}]{Ostrovskaya1999}%
  \BibitemOpen
  \bibfield  {author} {\bibinfo {author} {\bibfnamefont {E.~A.}\ \bibnamefont
  {Ostrovskaya}}, \bibinfo {author} {\bibfnamefont {Y.~S.}\ \bibnamefont
  {Kivshar}}, \bibinfo {author} {\bibfnamefont {Z.}~\bibnamefont {Chen}}, \
  and\ \bibinfo {author} {\bibfnamefont {M.}~\bibnamefont {Segev}},\ }\href
  {\doibase 10.1364/OL.24.000327} {\bibfield  {journal} {\bibinfo  {journal}
  {Opt. Lett.}\ }\textbf {\bibinfo {volume} {24}},\ \bibinfo {pages} {327}
  (\bibinfo {year} {1999})}\BibitemShut {NoStop}%
\bibitem [{\citenamefont {Inouye}\ \emph {et~al.}(1998)\citenamefont {Inouye},
  \citenamefont {Andrews}, \citenamefont {Stenger}, \citenamefont {Miesner},
  \citenamefont {{Stamper-Kurn}},\ and\ \citenamefont {Ketterle}}]{Inouye1998}%
  \BibitemOpen
  \bibfield  {author} {\bibinfo {author} {\bibfnamefont {S.}~\bibnamefont
  {Inouye}}, \bibinfo {author} {\bibfnamefont {M.~R.}\ \bibnamefont {Andrews}},
  \bibinfo {author} {\bibfnamefont {J.}~\bibnamefont {Stenger}}, \bibinfo
  {author} {\bibfnamefont {H.-J.}\ \bibnamefont {Miesner}}, \bibinfo {author}
  {\bibfnamefont {D.~M.}\ \bibnamefont {{Stamper-Kurn}}}, \ and\ \bibinfo
  {author} {\bibfnamefont {W.}~\bibnamefont {Ketterle}},\ }\href {\doibase
  10.1038/32354} {\bibfield  {journal} {\bibinfo  {journal} {Nature}\ }\textbf
  {\bibinfo {volume} {392}},\ \bibinfo {pages} {151} (\bibinfo {year}
  {1998})}\BibitemShut {NoStop}%
\bibitem [{\citenamefont {Chin}\ \emph {et~al.}(2010)\citenamefont {Chin},
  \citenamefont {Grimm}, \citenamefont {Julienne},\ and\ \citenamefont
  {Tiesinga}}]{Chin2010}%
  \BibitemOpen
  \bibfield  {author} {\bibinfo {author} {\bibfnamefont {C.}~\bibnamefont
  {Chin}}, \bibinfo {author} {\bibfnamefont {R.}~\bibnamefont {Grimm}},
  \bibinfo {author} {\bibfnamefont {P.}~\bibnamefont {Julienne}}, \ and\
  \bibinfo {author} {\bibfnamefont {E.}~\bibnamefont {Tiesinga}},\ }\href
  {\doibase 10.1103/RevModPhys.82.1225} {\bibfield  {journal} {\bibinfo
  {journal} {Rev. Mod. Phys.}\ }\textbf {\bibinfo {volume} {82}},\ \bibinfo
  {pages} {1225} (\bibinfo {year} {2010})}\BibitemShut {NoStop}%
\bibitem [{\citenamefont {Kr{\"o}kel}\ \emph {et~al.}(1988)\citenamefont
  {Kr{\"o}kel}, \citenamefont {Halas}, \citenamefont {Giuliani},\ and\
  \citenamefont {Grischkowsky}}]{Krokel1988}%
  \BibitemOpen
  \bibfield  {author} {\bibinfo {author} {\bibfnamefont {D.}~\bibnamefont
  {Kr{\"o}kel}}, \bibinfo {author} {\bibfnamefont {N.~J.}\ \bibnamefont
  {Halas}}, \bibinfo {author} {\bibfnamefont {G.}~\bibnamefont {Giuliani}}, \
  and\ \bibinfo {author} {\bibfnamefont {D.}~\bibnamefont {Grischkowsky}},\
  }\href {\doibase 10.1103/PhysRevLett.60.29} {\bibfield  {journal} {\bibinfo
  {journal} {Phys. Rev. Lett.}\ }\textbf {\bibinfo {volume} {60}},\ \bibinfo
  {pages} {29} (\bibinfo {year} {1988})}\BibitemShut {NoStop}%
\bibitem [{\citenamefont {Kamchatnov}\ \emph {et~al.}(2002)\citenamefont
  {Kamchatnov}, \citenamefont {Kraenkel},\ and\ \citenamefont
  {Umarov}}]{Kamchatnov2002}%
  \BibitemOpen
  \bibfield  {author} {\bibinfo {author} {\bibfnamefont {A.~M.}\ \bibnamefont
  {Kamchatnov}}, \bibinfo {author} {\bibfnamefont {R.~A.}\ \bibnamefont
  {Kraenkel}}, \ and\ \bibinfo {author} {\bibfnamefont {B.~A.}\ \bibnamefont
  {Umarov}},\ }\href {\doibase 10.1103/PhysRevE.66.036609} {\bibfield
  {journal} {\bibinfo  {journal} {Phys. Rev. E}\ }\textbf {\bibinfo {volume}
  {66}},\ \bibinfo {pages} {036609} (\bibinfo {year} {2002})}\BibitemShut
  {NoStop}%
\bibitem [{\citenamefont {Nikolov}\ \emph {et~al.}(2004)\citenamefont
  {Nikolov}, \citenamefont {Neshev}, \citenamefont {Kr{\'o}likowski},
  \citenamefont {Bang}, \citenamefont {Rasmussen},\ and\ \citenamefont
  {Christiansen}}]{Nikolov2004}%
  \BibitemOpen
  \bibfield  {author} {\bibinfo {author} {\bibfnamefont {N.~I.}\ \bibnamefont
  {Nikolov}}, \bibinfo {author} {\bibfnamefont {D.}~\bibnamefont {Neshev}},
  \bibinfo {author} {\bibfnamefont {W.}~\bibnamefont {Kr{\'o}likowski}},
  \bibinfo {author} {\bibfnamefont {O.}~\bibnamefont {Bang}}, \bibinfo {author}
  {\bibfnamefont {J.~J.}\ \bibnamefont {Rasmussen}}, \ and\ \bibinfo {author}
  {\bibfnamefont {P.~L.}\ \bibnamefont {Christiansen}},\ }\href {\doibase
  10.1364/OL.29.000286} {\bibfield  {journal} {\bibinfo  {journal} {Opt.
  Lett.}\ }\textbf {\bibinfo {volume} {29}},\ \bibinfo {pages} {286} (\bibinfo
  {year} {2004})}\BibitemShut {NoStop}%
\bibitem [{\citenamefont {{Dabrowska-W{\"u}ster}}\ \emph
  {et~al.}(2009)\citenamefont {{Dabrowska-W{\"u}ster}}, \citenamefont
  {W{\"u}ster},\ and\ \citenamefont {Davis}}]{Dabrowska-Wuster2009}%
  \BibitemOpen
  \bibfield  {author} {\bibinfo {author} {\bibfnamefont {B.~J.}\ \bibnamefont
  {{Dabrowska-W{\"u}ster}}}, \bibinfo {author} {\bibfnamefont {S.}~\bibnamefont
  {W{\"u}ster}}, \ and\ \bibinfo {author} {\bibfnamefont {M.~J.}\ \bibnamefont
  {Davis}},\ }\href {\doibase 10.1088/1367-2630/11/5/053017} {\bibfield
  {journal} {\bibinfo  {journal} {New J. Phys.}\ }\textbf {\bibinfo {volume}
  {11}},\ \bibinfo {pages} {053017} (\bibinfo {year} {2009})}\BibitemShut
  {NoStop}%
\bibitem [{\citenamefont {Prinari}\ \emph {et~al.}(2006)\citenamefont
  {Prinari}, \citenamefont {Ablowitz},\ and\ \citenamefont
  {Biondini}}]{Prinari2006}%
  \BibitemOpen
  \bibfield  {author} {\bibinfo {author} {\bibfnamefont {B.}~\bibnamefont
  {Prinari}}, \bibinfo {author} {\bibfnamefont {M.~J.}\ \bibnamefont
  {Ablowitz}}, \ and\ \bibinfo {author} {\bibfnamefont {G.}~\bibnamefont
  {Biondini}},\ }\href {\doibase 10.1063/1.2209169} {\bibfield  {journal}
  {\bibinfo  {journal} {J. Math. Phys.}\ }\textbf {\bibinfo {volume} {47}},\
  \bibinfo {pages} {063508} (\bibinfo {year} {2006})}\BibitemShut {NoStop}%
\bibitem [{\citenamefont {Biondini}\ and\ \citenamefont
  {Fagerstrom}(2015)}]{Biondini2015}%
  \BibitemOpen
  \bibfield  {author} {\bibinfo {author} {\bibfnamefont {G.}~\bibnamefont
  {Biondini}}\ and\ \bibinfo {author} {\bibfnamefont {E.}~\bibnamefont
  {Fagerstrom}},\ }\href {\doibase 10.1137/140965089} {\bibfield  {journal}
  {\bibinfo  {journal} {SIAM J. Appl. Math.}\ }\textbf {\bibinfo {volume}
  {75}},\ \bibinfo {pages} {136} (\bibinfo {year} {2015})}\BibitemShut
  {NoStop}%
\bibitem [{\citenamefont {Biondini}\ and\ \citenamefont
  {Kraus}(2015)}]{Biondini2015a}%
  \BibitemOpen
  \bibfield  {author} {\bibinfo {author} {\bibfnamefont {G.}~\bibnamefont
  {Biondini}}\ and\ \bibinfo {author} {\bibfnamefont {D.}~\bibnamefont
  {Kraus}},\ }\href {\doibase 10.1137/130943479} {\bibfield  {journal}
  {\bibinfo  {journal} {SIAM J. Math. Anal.}\ }\textbf {\bibinfo {volume}
  {47}},\ \bibinfo {pages} {706} (\bibinfo {year} {2015})}\BibitemShut
  {NoStop}%
\bibitem [{\citenamefont {Faddeev}\ and\ \citenamefont
  {Takhtajan}(2007)}]{Faddeev2007}%
  \BibitemOpen
  \bibfield  {author} {\bibinfo {author} {\bibfnamefont {L.}~\bibnamefont
  {Faddeev}}\ and\ \bibinfo {author} {\bibfnamefont {L.}~\bibnamefont
  {Takhtajan}},\ }\href@noop {} {\emph {\bibinfo {title} {Hamiltonian
  {{Methods}} in the {{Theory}} of {{Solitons}}}}}\ (\bibinfo  {publisher}
  {{Springer Science \& Business Media}},\ \bibinfo {year} {2007})\BibitemShut
  {NoStop}%
\bibitem [{\citenamefont {Bogoliubov}(1947)}]{Bogoliubov1947}%
  \BibitemOpen
  \bibfield  {author} {\bibinfo {author} {\bibfnamefont {N.}~\bibnamefont
  {Bogoliubov}},\ }\href@noop {} {\bibfield  {journal} {\bibinfo  {journal} {J.
  Phys.}\ }\textbf {\bibinfo {volume} {11}},\ \bibinfo {pages} {10} (\bibinfo
  {year} {1947})}\BibitemShut {NoStop}%
\bibitem [{\citenamefont {Lee}\ \emph {et~al.}(1957)\citenamefont {Lee},
  \citenamefont {Huang},\ and\ \citenamefont {Yang}}]{Lee1957}%
  \BibitemOpen
  \bibfield  {author} {\bibinfo {author} {\bibfnamefont {T.~D.}\ \bibnamefont
  {Lee}}, \bibinfo {author} {\bibfnamefont {K.}~\bibnamefont {Huang}}, \ and\
  \bibinfo {author} {\bibfnamefont {C.~N.}\ \bibnamefont {Yang}},\ }\href
  {\doibase 10.1103/PhysRev.106.1135} {\bibfield  {journal} {\bibinfo
  {journal} {Phys. Rev.}\ }\textbf {\bibinfo {volume} {106}},\ \bibinfo {pages}
  {1135} (\bibinfo {year} {1957})}\BibitemShut {NoStop}%
\bibitem [{\citenamefont {Dobrek}\ \emph {et~al.}(1999)\citenamefont {Dobrek},
  \citenamefont {Gajda}, \citenamefont {Lewenstein}, \citenamefont {Sengstock},
  \citenamefont {Birkl},\ and\ \citenamefont {Ertmer}}]{Dobrek1999}%
  \BibitemOpen
  \bibfield  {author} {\bibinfo {author} {\bibfnamefont {{\L}.}~\bibnamefont
  {Dobrek}}, \bibinfo {author} {\bibfnamefont {M.}~\bibnamefont {Gajda}},
  \bibinfo {author} {\bibfnamefont {M.}~\bibnamefont {Lewenstein}}, \bibinfo
  {author} {\bibfnamefont {K.}~\bibnamefont {Sengstock}}, \bibinfo {author}
  {\bibfnamefont {G.}~\bibnamefont {Birkl}}, \ and\ \bibinfo {author}
  {\bibfnamefont {W.}~\bibnamefont {Ertmer}},\ }\href {\doibase
  10.1103/PhysRevA.60.R3381} {\bibfield  {journal} {\bibinfo  {journal} {Phys.
  Rev. A}\ }\textbf {\bibinfo {volume} {60}},\ \bibinfo {pages} {R3381}
  (\bibinfo {year} {1999})}\BibitemShut {NoStop}%
\bibitem [{\citenamefont {Wu}\ \emph {et~al.}(2002)\citenamefont {Wu},
  \citenamefont {Liu},\ and\ \citenamefont {Niu}}]{Wu2002}%
  \BibitemOpen
  \bibfield  {author} {\bibinfo {author} {\bibfnamefont {B.}~\bibnamefont
  {Wu}}, \bibinfo {author} {\bibfnamefont {J.}~\bibnamefont {Liu}}, \ and\
  \bibinfo {author} {\bibfnamefont {Q.}~\bibnamefont {Niu}},\ }\href {\doibase
  10.1103/PhysRevLett.88.034101} {\bibfield  {journal} {\bibinfo  {journal}
  {Phys. Rev. Lett.}\ }\textbf {\bibinfo {volume} {88}},\ \bibinfo {pages}
  {034101} (\bibinfo {year} {2002})}\BibitemShut {NoStop}%
\bibitem [{\citenamefont {Fritsch}\ \emph {et~al.}(2020)\citenamefont
  {Fritsch}, \citenamefont {Lu}, \citenamefont {Reid}, \citenamefont
  {Pi{\~n}eiro},\ and\ \citenamefont {Spielman}}]{Fritsch2020}%
  \BibitemOpen
  \bibfield  {author} {\bibinfo {author} {\bibfnamefont {A.~R.}\ \bibnamefont
  {Fritsch}}, \bibinfo {author} {\bibfnamefont {M.}~\bibnamefont {Lu}},
  \bibinfo {author} {\bibfnamefont {G.~H.}\ \bibnamefont {Reid}}, \bibinfo
  {author} {\bibfnamefont {A.~M.}\ \bibnamefont {Pi{\~n}eiro}}, \ and\ \bibinfo
  {author} {\bibfnamefont {I.~B.}\ \bibnamefont {Spielman}},\ }\href {\doibase
  10.1103/PhysRevA.101.053629} {\bibfield  {journal} {\bibinfo  {journal}
  {Phys. Rev. A}\ }\textbf {\bibinfo {volume} {101}},\ \bibinfo {pages}
  {053629} (\bibinfo {year} {2020})}\BibitemShut {NoStop}%
\bibitem [{\citenamefont {El}\ and\ \citenamefont {Hoefer}(2016)}]{El2016}%
  \BibitemOpen
  \bibfield  {author} {\bibinfo {author} {\bibfnamefont {G.~A.}\ \bibnamefont
  {El}}\ and\ \bibinfo {author} {\bibfnamefont {M.~A.}\ \bibnamefont
  {Hoefer}},\ }\href {\doibase 10.1016/j.physd.2016.04.006} {\bibfield
  {journal} {\bibinfo  {journal} {Physica D: Nonlinear Phenomena}\ }\bibinfo
  {series} {Dispersive {{Hydrodynamics}}},\ \textbf {\bibinfo {volume} {333}},\
  \bibinfo {pages} {11} (\bibinfo {year} {2016})}\BibitemShut {NoStop}%
\bibitem [{\citenamefont {Kodama}\ and\ \citenamefont
  {Wabnitz}(1995)}]{Kodama1995}%
  \BibitemOpen
  \bibfield  {author} {\bibinfo {author} {\bibfnamefont {Y.}~\bibnamefont
  {Kodama}}\ and\ \bibinfo {author} {\bibfnamefont {S.}~\bibnamefont
  {Wabnitz}},\ }\href {\doibase 10.1364/OL.20.002291} {\bibfield  {journal}
  {\bibinfo  {journal} {Opt. Lett.}\ }\textbf {\bibinfo {volume} {20}},\
  \bibinfo {pages} {2291} (\bibinfo {year} {1995})}\BibitemShut {NoStop}%
\bibitem [{\citenamefont {Whitham}(1974)}]{Whitham1974}%
  \BibitemOpen
  \bibfield  {author} {\bibinfo {author} {\bibfnamefont {G.~B.}\ \bibnamefont
  {Whitham}},\ }\href@noop {} {\emph {\bibinfo {title} {Linear and {{Nonlinear
  Waves}}}}}\ (\bibinfo  {publisher} {{John Wiley \& Sons}},\ \bibinfo {year}
  {1974})\BibitemShut {NoStop}%
\bibitem [{\citenamefont {Pavlov}(1987)}]{Pavlov1987}%
  \BibitemOpen
  \bibfield  {author} {\bibinfo {author} {\bibfnamefont {M.~V.}\ \bibnamefont
  {Pavlov}},\ }\href {\doibase 10.1007/BF01017090} {\bibfield  {journal}
  {\bibinfo  {journal} {Theor. Math. Phys.}\ }\textbf {\bibinfo {volume}
  {71}},\ \bibinfo {pages} {584} (\bibinfo {year} {1987})}\BibitemShut
  {NoStop}%
\bibitem [{\citenamefont {Gurevich}\ and\ \citenamefont
  {Krylov}(1987)}]{Gurevich1987}%
  \BibitemOpen
  \bibfield  {author} {\bibinfo {author} {\bibfnamefont {V.}~\bibnamefont
  {Gurevich}}\ and\ \bibinfo {author} {\bibfnamefont {A.~L.}\ \bibnamefont
  {Krylov}},\ }\href@noop {} {\bibfield  {journal} {\bibinfo  {journal} {JETP}\
  }\textbf {\bibinfo {volume} {65}},\ \bibinfo {pages} {944} (\bibinfo {year}
  {1987})}\BibitemShut {NoStop}%
\bibitem [{\citenamefont {El}\ \emph {et~al.}(1995)\citenamefont {El},
  \citenamefont {Geogjaev}, \citenamefont {Gurevich},\ and\ \citenamefont
  {Krylov}}]{El1995}%
  \BibitemOpen
  \bibfield  {author} {\bibinfo {author} {\bibfnamefont {G.~A.}\ \bibnamefont
  {El}}, \bibinfo {author} {\bibfnamefont {V.~V.}\ \bibnamefont {Geogjaev}},
  \bibinfo {author} {\bibfnamefont {A.~V.}\ \bibnamefont {Gurevich}}, \ and\
  \bibinfo {author} {\bibfnamefont {A.~L.}\ \bibnamefont {Krylov}},\ }\href
  {\doibase 10.1016/0167-2789(95)00147-V} {\bibfield  {journal} {\bibinfo
  {journal} {Physica D: Nonlinear Phenomena}\ }\bibinfo {series} {Proceedings
  of the {{Conference}} on {{The Nonlinear Schrodinger Equation}}},\ \textbf
  {\bibinfo {volume} {87}},\ \bibinfo {pages} {186} (\bibinfo {year}
  {1995})}\BibitemShut {NoStop}%
\bibitem [{\citenamefont {Hoefer}\ \emph {et~al.}(2006)\citenamefont {Hoefer},
  \citenamefont {Ablowitz}, \citenamefont {Coddington}, \citenamefont
  {Cornell}, \citenamefont {Engels},\ and\ \citenamefont
  {Schweikhard}}]{Hoefer2006}%
  \BibitemOpen
  \bibfield  {author} {\bibinfo {author} {\bibfnamefont {M.~A.}\ \bibnamefont
  {Hoefer}}, \bibinfo {author} {\bibfnamefont {M.~J.}\ \bibnamefont
  {Ablowitz}}, \bibinfo {author} {\bibfnamefont {I.}~\bibnamefont
  {Coddington}}, \bibinfo {author} {\bibfnamefont {E.~A.}\ \bibnamefont
  {Cornell}}, \bibinfo {author} {\bibfnamefont {P.}~\bibnamefont {Engels}}, \
  and\ \bibinfo {author} {\bibfnamefont {V.}~\bibnamefont {Schweikhard}},\
  }\href {\doibase 10.1103/PhysRevA.74.023623} {\bibfield  {journal} {\bibinfo
  {journal} {Phys. Rev. A}\ }\textbf {\bibinfo {volume} {74}},\ \bibinfo
  {pages} {023623} (\bibinfo {year} {2006})}\BibitemShut {NoStop}%
\bibitem [{\citenamefont {Hoefer}\ \emph {et~al.}(2008)\citenamefont {Hoefer},
  \citenamefont {Ablowitz},\ and\ \citenamefont {Engels}}]{Hoefer2008}%
  \BibitemOpen
  \bibfield  {author} {\bibinfo {author} {\bibfnamefont {M.~A.}\ \bibnamefont
  {Hoefer}}, \bibinfo {author} {\bibfnamefont {M.~J.}\ \bibnamefont
  {Ablowitz}}, \ and\ \bibinfo {author} {\bibfnamefont {P.}~\bibnamefont
  {Engels}},\ }\href {\doibase 10.1103/PhysRevLett.100.084504} {\bibfield
  {journal} {\bibinfo  {journal} {Phys. Rev. Lett.}\ }\textbf {\bibinfo
  {volume} {100}},\ \bibinfo {pages} {084504} (\bibinfo {year}
  {2008})}\BibitemShut {NoStop}%
\bibitem [{\citenamefont {Biondini}\ and\ \citenamefont
  {Kodama}(2006)}]{Biondini2006}%
  \BibitemOpen
  \bibfield  {author} {\bibinfo {author} {\bibfnamefont {G.}~\bibnamefont
  {Biondini}}\ and\ \bibinfo {author} {\bibfnamefont {Y.}~\bibnamefont
  {Kodama}},\ }\href {\doibase 10.1007/s00332-005-0733-2} {\bibfield  {journal}
  {\bibinfo  {journal} {J Nonlinear Sci}\ }\textbf {\bibinfo {volume} {16}},\
  \bibinfo {pages} {435} (\bibinfo {year} {2006})}\BibitemShut {NoStop}%
\bibitem [{\citenamefont {Hoefer}\ and\ \citenamefont
  {Ablowitz}(2007)}]{Hoefer2007}%
  \BibitemOpen
  \bibfield  {author} {\bibinfo {author} {\bibfnamefont {M.~A.}\ \bibnamefont
  {Hoefer}}\ and\ \bibinfo {author} {\bibfnamefont {M.~J.}\ \bibnamefont
  {Ablowitz}},\ }\href {\doibase 10.1016/j.physd.2007.07.017} {\bibfield
  {journal} {\bibinfo  {journal} {Physica D: Nonlinear Phenomena}\ }\textbf
  {\bibinfo {volume} {236}},\ \bibinfo {pages} {44} (\bibinfo {year}
  {2007})}\BibitemShut {NoStop}%
\bibitem [{\citenamefont {Forest}\ and\ \citenamefont
  {Lee}(1986)}]{Forest1986}%
  \BibitemOpen
  \bibfield  {author} {\bibinfo {author} {\bibfnamefont {M.~G.}\ \bibnamefont
  {Forest}}\ and\ \bibinfo {author} {\bibfnamefont {J.-E.}\ \bibnamefont
  {Lee}},\ }in\ \href {\doibase 10.1007/978-1-4613-8689-6_3} {\emph {\bibinfo
  {booktitle} {Oscillation {{Theory}}, {{Computation}}, and {{Methods}} of
  {{Compensated Compactness}}}}},\ \bibinfo {series and number} {The {{IMA
  Volumes}} in {{Mathematics}} and {{Its Applications}}},\ \bibinfo {editor}
  {edited by\ \bibinfo {editor} {\bibfnamefont {C.}~\bibnamefont {Dafermos}},
  \bibinfo {editor} {\bibfnamefont {J.~L.}\ \bibnamefont {Ericksen}}, \bibinfo
  {editor} {\bibfnamefont {D.}~\bibnamefont {Kinderlehrer}}, \ and\ \bibinfo
  {editor} {\bibfnamefont {M.}~\bibnamefont {Slemrod}}}\ (\bibinfo  {publisher}
  {{Springer}},\ \bibinfo {address} {{New York, NY}},\ \bibinfo {year} {1986})\
  pp.\ \bibinfo {pages} {35--69}\BibitemShut {NoStop}%
\bibitem [{\citenamefont {Katsimiga}\ \emph
  {et~al.}(2018{\natexlab{a}})\citenamefont {Katsimiga}, \citenamefont
  {Mistakidis}, \citenamefont {Koutentakis}, \citenamefont {Kevrekidis},\ and\
  \citenamefont {Schmelcher}}]{Katsimiga2018}%
  \BibitemOpen
  \bibfield  {author} {\bibinfo {author} {\bibfnamefont {G.~C.}\ \bibnamefont
  {Katsimiga}}, \bibinfo {author} {\bibfnamefont {S.~I.}\ \bibnamefont
  {Mistakidis}}, \bibinfo {author} {\bibfnamefont {G.~M.}\ \bibnamefont
  {Koutentakis}}, \bibinfo {author} {\bibfnamefont {P.~G.}\ \bibnamefont
  {Kevrekidis}}, \ and\ \bibinfo {author} {\bibfnamefont {P.}~\bibnamefont
  {Schmelcher}},\ }\href {\doibase 10.1103/PhysRevA.98.013632} {\bibfield
  {journal} {\bibinfo  {journal} {Phys. Rev. A}\ }\textbf {\bibinfo {volume}
  {98}},\ \bibinfo {pages} {013632} (\bibinfo {year}
  {2018}{\natexlab{a}})}\BibitemShut {NoStop}%
\bibitem [{\citenamefont {Katsimiga}\ \emph {et~al.}(2020)\citenamefont
  {Katsimiga}, \citenamefont {Mistakidis}, \citenamefont {Bersano},
  \citenamefont {Ome}, \citenamefont {Mossman}, \citenamefont {Mukherjee},
  \citenamefont {Schmelcher}, \citenamefont {Engels},\ and\ \citenamefont
  {Kevrekidis}}]{Katsimiga2020}%
  \BibitemOpen
  \bibfield  {author} {\bibinfo {author} {\bibfnamefont {G.~C.}\ \bibnamefont
  {Katsimiga}}, \bibinfo {author} {\bibfnamefont {S.~I.}\ \bibnamefont
  {Mistakidis}}, \bibinfo {author} {\bibfnamefont {T.~M.}\ \bibnamefont
  {Bersano}}, \bibinfo {author} {\bibfnamefont {M.~K.~H.}\ \bibnamefont {Ome}},
  \bibinfo {author} {\bibfnamefont {S.~M.}\ \bibnamefont {Mossman}}, \bibinfo
  {author} {\bibfnamefont {K.}~\bibnamefont {Mukherjee}}, \bibinfo {author}
  {\bibfnamefont {P.}~\bibnamefont {Schmelcher}}, \bibinfo {author}
  {\bibfnamefont {P.}~\bibnamefont {Engels}}, \ and\ \bibinfo {author}
  {\bibfnamefont {P.~G.}\ \bibnamefont {Kevrekidis}},\ }\href {\doibase
  10.1103/PhysRevA.102.023301} {\bibfield  {journal} {\bibinfo  {journal}
  {Phys. Rev. A}\ }\textbf {\bibinfo {volume} {102}},\ \bibinfo {pages}
  {023301} (\bibinfo {year} {2020})}\BibitemShut {NoStop}%
\bibitem [{\citenamefont {Busch}\ and\ \citenamefont
  {Anglin}(2000)}]{Busch2000}%
  \BibitemOpen
  \bibfield  {author} {\bibinfo {author} {\bibfnamefont {T.}~\bibnamefont
  {Busch}}\ and\ \bibinfo {author} {\bibfnamefont {J.~R.}\ \bibnamefont
  {Anglin}},\ }\href {\doibase 10.1103/PhysRevLett.84.2298} {\bibfield
  {journal} {\bibinfo  {journal} {Phys. Rev. Lett.}\ }\textbf {\bibinfo
  {volume} {84}},\ \bibinfo {pages} {2298} (\bibinfo {year}
  {2000})}\BibitemShut {NoStop}%
\bibitem [{\citenamefont {Katsimiga}\ \emph
  {et~al.}(2018{\natexlab{b}})\citenamefont {Katsimiga}, \citenamefont
  {Kevrekidis}, \citenamefont {Prinari}, \citenamefont {Biondini},\ and\
  \citenamefont {Schmelcher}}]{Katsimiga2018a}%
  \BibitemOpen
  \bibfield  {author} {\bibinfo {author} {\bibfnamefont {G.~C.}\ \bibnamefont
  {Katsimiga}}, \bibinfo {author} {\bibfnamefont {P.~G.}\ \bibnamefont
  {Kevrekidis}}, \bibinfo {author} {\bibfnamefont {B.}~\bibnamefont {Prinari}},
  \bibinfo {author} {\bibfnamefont {G.}~\bibnamefont {Biondini}}, \ and\
  \bibinfo {author} {\bibfnamefont {P.}~\bibnamefont {Schmelcher}},\ }\href
  {\doibase 10.1103/PhysRevA.97.043623} {\bibfield  {journal} {\bibinfo
  {journal} {Phys. Rev. A}\ }\textbf {\bibinfo {volume} {97}},\ \bibinfo
  {pages} {043623} (\bibinfo {year} {2018}{\natexlab{b}})}\BibitemShut
  {NoStop}%
\bibitem [{\citenamefont {Scherer}\ \emph {et~al.}(2007)\citenamefont
  {Scherer}, \citenamefont {Weiler}, \citenamefont {Neely},\ and\ \citenamefont
  {Anderson}}]{Scherer2007}%
  \BibitemOpen
  \bibfield  {author} {\bibinfo {author} {\bibfnamefont {D.~R.}\ \bibnamefont
  {Scherer}}, \bibinfo {author} {\bibfnamefont {C.~N.}\ \bibnamefont {Weiler}},
  \bibinfo {author} {\bibfnamefont {T.~W.}\ \bibnamefont {Neely}}, \ and\
  \bibinfo {author} {\bibfnamefont {B.~P.}\ \bibnamefont {Anderson}},\ }\href
  {\doibase 10.1103/PhysRevLett.98.110402} {\bibfield  {journal} {\bibinfo
  {journal} {Phys. Rev. Lett.}\ }\textbf {\bibinfo {volume} {98}},\ \bibinfo
  {pages} {110402} (\bibinfo {year} {2007})}\BibitemShut {NoStop}%
\end{thebibliography}%

\end{document}